# Beamforming and Rate Allocation in MISO Cognitive Radio Networks

Ali Tajer[*]    Narayan Prasad [†]    Xiaodong Wang[*]

**Abstract**

We consider decentralized multi-antenna cognitive radio networks where secondary (cognitive) users are granted simultaneous spectrum access along with license-holding (primary) users. We treat the problem of *distributed* beamforming and rate allocation for the secondary users such that the minimum weighted secondary rate is maximized. Such an optimization is subject to (1) a limited weighted sum-power budget for the secondary users and (2) guaranteed protection for the primary users in the sense that the interference level imposed on each primary receiver does not exceed a specified level. Based on the decoding method deployed by the secondary receivers, we consider three scenarios for solving this problem. In the first scenario each secondary receiver decodes only its designated transmitter while suppressing the rest as Gaussian interferers (single-user decoding). In the second case each secondary receiver employs the maximum likelihood decoder (MLD) to jointly decode all secondary transmissions, and in the third one each secondary receiver uses the unconstrained group decoder (UGD). By deploying the UGD, each secondary user is allowed to decode any arbitrary subset of users (which contains its designated user) after suppressing or canceling the remaining users.

## 1 Introduction

In classical cognitive radio systems the secondary users can only transmit in white spaces which denote the frequency bands (or time intervals) that the primary (or licensed) users are silent [1]. On the other hand, in generalized cognitive radio systems, the secondary users can also transmit simultaneously with primary users, as long as certain co-existence constraints are satisfied [2, 3].

[*]Electrical Engineering Department, Columbia University, New York, NY 10027 (email: {tajer, wangx}@ee.columbia.edu).
[†]NEC Labs America, Princeton, NJ 08540 (email: prasad@nec-labs.com)

Clearly the latter systems can achieve higher spectral efficiencies but at the expense of additional side-information at the secondary users and increased signaling overhead. We consider decentralized multi-antenna cognitive radio networks where secondary transceivers can co-exist with primary ones. However, in our setup no secondary transmitter has access to any primary user's codebook. Instead, each secondary transmitter employs beamforming to communicate with its desired receiver while ensuring that the aggregate interference seen by each primary receiver does not exceed a specified level (interference margin).

Our goal is to design optimal beamformers for the secondary users and assign rates to them in a distributed fashion, in order to maximize the smallest weighted rate among secondary users. This optimization is subject to a weighted sum-power constraint on the secondary users as well as the interference margin constraints imposed by the primary users. Based on the decoding scheme deployed by the secondary users, we consider three scenarios for solving this problem. In the first one each secondary receiver employs a minimum mean-squared error (MMSE) decoder and attempts to decode only the signal transmitted by its designated transmitter after suppressing the remaining signals via linear filtering (a.k.a. single user decoding). We propose an efficient distributed algorithm that yields a globally optimal set of beamformers. The resulting beamformers are optimal for networks which do not support exchange of codebooks among secondary transceivers or where more advanced decoding at the secondary receivers is not feasible due to complexity constraints. In this context we note that [4] has optimally solved the linear transmit beamforming problem in a MISO cognitive radio network [1] with a single secondary transceiver. A variation of the beamforming problem has also been optimally solved in [5] for a MISO cognitive radio network with a single secondary and a single primary transceiver but where the channel vector from the secondary transmitter to the primary receiver is imperfectly known to the secondary user. Other related works include [6] which is a comprehensive work on precoder design in MISO point-to-multi-point channels and [7] which proposes a distributed beamforming algorithm to minimize the weighted sum power subject to SINR constraints in the downlink MISO multi-cell network.

In the second scenario, we assume that each secondary receiver is equipped with a maximum likelihood decoder (MLD) which is used to jointly decode all secondary transmissions. We show that in this scenario, the beamforming design problem can be posed as a non-convex optimization problem with a quadratic objective and *indefinite* quadratic constraints. Solving such a problem optimally in its general form is intractable [8]. Nevertheless, after a convex relaxation it can be converted to a semi-definite program (SDP). This design procedure for MLD requires central processing and does not seem practical for decentralized cognitive networks.

---

[1] A cognitive radio network where all the receivers have a single receive antenna is referred to here as a MISO cognitive radio network.

In the third scenario, we assume that each secondary receiver uses the unconstrained group decoder (UGD) [9] and is allowed to jointly decode any subset of secondary users containing its desired user, after suppressing or canceling the rest of users. We provide an explicit formulation of the optimization problem for this scenario. The resulting problem is non-convex and hence cannot be efficiently solved even in a centralized setup.

As a remedy for the MLD and UGD, we adopt a two-step sub-optimal approach. In the first step, we obtain a set of beamformers which is optimal under MMSE decoding at each receiver. In the second step, in order to boost the spectral efficiency we exploit the fact that the secondary users by employing more advanced decoders can support higher data rates. Therefore, we propose allocating excess rates to the secondary users beyond their minimum acceptable rates, such that (1) *weighted max-min* fairness is maintained, (2) all the secondary users remain decodable at their respective receivers and (3) the rate assignment is pareto-optimal. A key feature of our proposed distributed rate allocation algorithms is that the complexity at each secondary receiver is only polynomial in the number of secondary users. In this context, we note that efficient rate allocation in a single-antenna interference channel (IC) with single-codebook and fixed power per-user has been recently investigated in [10, 11]. In particular, [11] considers a $K$-user IC where each user employs the successive interference cancelation (SIC) based decoder and obtains a max-min fair decentralized rate allocation algorithm. [10] considers an $K$-user Gaussian IC and solves the problem of maximizing the desired user's rate at a particular receiver given the transmission rates of the other users. In addition, [10] also proposes sequential and iterative rate allocation algorithms which yield pareto-optimal rate-vectors albeit without a fairness guarantee.

The remainder of the paper is organized as follows. The model of the multi-antenna cognitive network is provided in Section 2 and the statements of the beamforming and rate optimization problems are formalized in Section 3. The design of the beamformers for the MMSE, ML and UGD receivers are discussed in Section 4 where we also provide a distributed algorithm for optimally solving the problem for MMSE receivers. Leveraging this beamforming design procedure, in Section 5 we consider the MLD and the UGD and provide distributed weighted max-min fair rate allocation algorithms for further boosting the rates of the secondary users. The simulation results are provided in Section 6 and Section 7 concludes the paper. To enhance the flow of the paper most of the proofs are confined in the appendices.

## 2 System Model

We consider a decentralized cognitive network comprising of $M_s$ secondary transmitter-receiver pairs co-existing with $M_p$ primary transceiver pairs via concurrent spectrum access. The secondary transceivers form a multi-antenna Gaussian interference channel (GIC) where $M_s$ transmitters each equipped with $N_s$ transmit antennas communicate with their designated single-antenna receivers. The primary transmitters and receivers have $N_p$ and 1 transmit and receive antennas, respectively. We assume quasi-static flat fading channels and denote the channels from the $j^{th}$ secondary transmitter to the $i^{th}$ secondary and primary receivers by $\boldsymbol{h}_{i,j}^{s,s} \in \mathbb{C}^{1 \times N_s}$ and $\boldsymbol{h}_{i,j}^{p,s} \in \mathbb{C}^{1 \times N_s}$, respectively, and denote the channels from the $j^{th}$ primary transmitter to the $i^{th}$ secondary and primary receivers by $\boldsymbol{h}_{i,j}^{s,p} \in \mathbb{C}^{1 \times N_p}$ and $\boldsymbol{h}_{i,j}^{p,p} \in \mathbb{C}^{1 \times N_p}$, respectively.

Let $x_i^s$ and $x_i^p$ be scalar complex-valued random variables with unit power, i.e., $\mathbb{E}[|x_i^s|^2] = \mathbb{E}[|x_i^p|^2] = 1$, representing the information symbols of the $i^{th}$ secondary and primary transmitters, respectively and let $\boldsymbol{w}_i^s \in \mathbb{C}^{N_s \times 1}$ and $\boldsymbol{w}_i^p \in \mathbb{C}^{N_p \times 1}$ denote their respective beamforming vectors. The received signals at the $i^{th}$ secondary and primary receivers are given by

$$y_i^s \triangleq \sum_{j=1}^{M_s} \boldsymbol{h}_{i,j}^{s,s} \boldsymbol{w}_j^s x_j^s + \sum_{j=1}^{M_p} \boldsymbol{h}_{i,j}^{s,p} \boldsymbol{w}_j^p x_j^p + z_i^s, \quad \text{for } i = 1, \ldots, M_s, \tag{1}$$

$$\text{and} \quad y_i^p \triangleq \sum_{j=1}^{M_s} \boldsymbol{h}_{i,j}^{p,s} \boldsymbol{w}_j^s x_j^s + \sum_{j=1}^{M_p} \boldsymbol{h}_{i,j}^{p,p} \boldsymbol{w}_j^p x_j^p + z_i^p, \quad \text{for } i = 1, \ldots, M_p, \tag{2}$$

where $z_i^s, z_i^p \in \mathbb{C}$ are the additive white Gaussian noise terms with variances $\sigma_i^s$ and $\sigma_i^p$, respectively. No primary (secondary) receiver tries to decode the signal intended for any secondary (primary) user.

## 3 Problem Statement

We denote the rate assigned to the $i^{th}$ secondary user by $R_i$ and we will say that for the given channel realization, choice of transmit beamformers and decoders employed by the secondary receivers, the rate vector $\boldsymbol{R} \triangleq [R_1, \ldots, R_{M_s}]$ is decodable (strictly decodable) if for any rate vector $\tilde{\boldsymbol{R}} \prec \boldsymbol{R}$ ($\tilde{\boldsymbol{R}} \preceq \boldsymbol{R}$) and any arbitrarily fixed $\epsilon > 0$, there exists a set of $M_s$ codes such that each secondary receiver can decode its desired user (secondary transmitter) with a probability of error no greater than $\epsilon$. The interference level seen by the $i^{th}$ primary receiver due to secondary transmissions is denoted by $J_i$ and is given by

$$J_i \triangleq \sum_{j=1}^{M_s} |\boldsymbol{h}_{i,j}^{p,s} \boldsymbol{w}_j^s|^2, \quad \text{for } i = 1, \ldots, M_p. \tag{3}$$

The $i^{th}$ primary receiver specifies a parameter $\beta_i$ which is the maximum interference it can tolerate from secondary transmissions. Let $\boldsymbol{J} \triangleq [J_1, \ldots, J_{M_p}]$ and define the interference margin vector $\boldsymbol{\beta} \triangleq [\beta_1, \ldots, \beta_{M_p}]$. We are interested in solving the following rate optimization problem.

For the given set of channel coefficients, choice of primary transmit beamformers, and decoders employed by the secondary receivers, we seek to maximize the worst-case secondary weighted rate such that the secondary weighted sum-power is below $P_0$ and the interference seen by the $i^{th}$ primary receiver does not exceed $\beta_i$, i.e.,

$$\mathcal{R}(P_0) = \begin{cases} \max_{\boldsymbol{R}, \{\boldsymbol{w}_i^s\}} & \min_i \frac{R_i}{\rho_i}, \\ \text{s.t.} & \sum_{i=1}^{M_s} \alpha_i \|\boldsymbol{w}_i^s\|^2 \leq P_0, \\ & \boldsymbol{J} \preceq \boldsymbol{\beta}, \\ & \boldsymbol{R} \text{ is decodable}, \end{cases} \tag{4}$$

where $\{\rho_i\}_{i=1}^{M_s}$ and $\{\alpha_i\}_{i=1}^{M_s}$ are all positive and account for weighting the individual rates and powers of the secondary users, respectively. Note that by definition, if $\boldsymbol{R}$ is decodable then any rate vector $\tilde{\boldsymbol{R}}$, where $\boldsymbol{0} \preceq \tilde{\boldsymbol{R}} \preceq \boldsymbol{R}$, is also decodable. Furthermore, the optimization problem in (4) is always feasible. We show that solving the rate optimization problem corresponding to $\mathcal{R}(P_0)$ can often be facilitated by alternatively solving a power optimization problem given by

$$\mathcal{P}(\boldsymbol{\rho}) = \begin{cases} \min_{\{\boldsymbol{w}_i^s\}} & \sum_{i=1}^{M_s} \alpha_i \|\boldsymbol{w}_i^s\|^2, \\ \text{s.t.} & \boldsymbol{J} \preceq \boldsymbol{\beta}, \\ & \text{Rate } \boldsymbol{\rho} \text{ is decodable}, \end{cases} \tag{5}$$

where $\boldsymbol{\rho} \triangleq [\rho_1, \ldots, \rho_{M_s}]$. Note that the optimization problem in (5) is not necessarily always feasible and we adopt the convention $\mathcal{P}(\boldsymbol{\rho}) = \infty$ when the problem is infeasible. It can be readily verified that $\mathcal{R}(P_0)$ is continuous and non-decreasing in $P_0$. Moreover, $\mathcal{P}(\boldsymbol{\rho} \cdot \rho_0)$ is also continuous and increasing in $\rho_0$ at any strictly feasible rate $\rho_0 \cdot \boldsymbol{\rho}$, i.e., at any strictly decodable rate vector $\rho_0 \cdot \boldsymbol{\rho}$ for which the margin constraints in (5) are satisfied with strict inequality. The rate and power optimization problems are related as follows.

**Theorem 1** *For any arbitrary set of $\boldsymbol{\alpha}, \boldsymbol{\rho}, \boldsymbol{\beta}$, the rate and power optimization problems corresponding to $\mathcal{R}(P_0)$ and $\mathcal{P}(\boldsymbol{\rho})$ are related as follows*

$$\mathcal{P}\left(\mathcal{R}(P_0) \cdot \boldsymbol{\rho}\right) \leq P_0, \tag{6}$$

*and if (5) is feasible then*

$$\mathcal{R}\left(\mathcal{P}(\boldsymbol{\rho})\right) = 1. $$

*Proof:* See Appendix A. ∎

Thus, the rate and power optimization problems can be deemed as complementary. It is noteworthy that the inequality in (6) becomes equality only if the weighted sum-power constraint in the rate optimization problem corresponding to $\mathcal{R}(P_0)$ holds with equality at each one of its optimal solutions. Also, for each choice of decoders considered in the sequel, strict inequality holds in (6) only if at-least one of the margin constraints in the power optimization problem corresponding to $\mathcal{P}\left(\mathcal{R}(P_0) \cdot \boldsymbol{\rho}\right)$ is active at each of its optimal solutions. This follows from the fact that for any given set of beamformers, wherein all secondary beamforming vectors are non-zero, we can strictly increase the rates of all secondary users by scaling their powers identically.

## 4 Beamformer Design

### 4.1 MMSE Receivers

In this section, we assume that each secondary receiver uses the MMSE single-user decoder which only decodes its desired user and treats the other users as Gaussian interferers. Then, for a given set of channel coefficients and choice of beamformers, the rate that can be achieved for the $i^{th}$ user is $R_i = \log(1 + \mathsf{SINR}_i)$, where $\mathsf{SINR}_i$ denotes the signal-to-interference-plus-noise ratio at the $i^{th}$ secondary receiver and is given by

$$\mathsf{SINR}_i \triangleq \frac{|\boldsymbol{h}_{i,i}^{s,s} \boldsymbol{w}_i^s|^2}{\sum_{j \neq i} |\boldsymbol{h}_{i,j}^{s,s} \boldsymbol{w}_j^s|^2 + \sum_j |\boldsymbol{h}_{i,j}^{s,p} \boldsymbol{w}_j^p|^2 + \sigma_i^s}. \tag{7}$$

We first provide a distributed algorithm for solving $\mathcal{P}(\boldsymbol{\rho})$. Then, by exploiting the connection between the problems $\mathcal{R}(P_0)$ and $\mathcal{P}(\boldsymbol{\rho})$, as established in Theorem 1, we use this algorithm to obtain another distributed algorithm for solving $\mathcal{R}(P_0)$. By defining $\gamma_i \triangleq 2^{\rho_i} - 1$ and $\boldsymbol{\gamma} \triangleq [\gamma_1, \ldots, \gamma_{M_s}]$ the optimization problem in (5) becomes equivalent to the problem

$$\tilde{\mathcal{P}}(\boldsymbol{\gamma}) = \begin{cases} \min_{\{\boldsymbol{w}_i^s\}} & \sum_{i=1}^{M_s} \alpha_i \|\boldsymbol{w}_i^s\|^2, \\ \text{s.t.} & \mathsf{SINR}_i \geq \gamma_i, \quad \text{for } i = 1, \ldots, M_s, \\ & \boldsymbol{J} \preceq \boldsymbol{\beta}. \end{cases} \qquad (8)$$

For solving the problem $\tilde{\mathcal{P}}(\boldsymbol{\gamma})$ we start off by investigating the feasibility of this problem. Next, we show that the problem can be solved efficiently as a second-order cone program (SOCP) and then propose a distributed algorithm that finds its globally optimal solution while abstaining from requiring a central controller.

For any set of channel realizations $\boldsymbol{h}_{i,j}^{s,s}$, $\boldsymbol{h}_{i,j}^{s,p}$, $\boldsymbol{h}_{i,j}^{p,s}$ & $\boldsymbol{h}_{i,j}^{p,p}$, and given $\boldsymbol{\alpha}, \boldsymbol{\gamma}, \boldsymbol{\beta}$ we define

$$\tilde{\boldsymbol{w}}_i^s \triangleq \sqrt{\alpha_i}\boldsymbol{w}_i^s \quad \text{and} \quad \tilde{\boldsymbol{h}}_{i,i}^{s,s} \triangleq \frac{\boldsymbol{h}_{i,i}^{s,s}}{\sqrt{\alpha_i \gamma_i}} \quad \text{for } i = 1, \ldots, M_s,$$

$$\tilde{\boldsymbol{h}}_{i,j}^{p,s} \triangleq \frac{\boldsymbol{h}_{i,j}^{p,s}}{\sqrt{\beta_i \alpha_j}} \quad \text{for } i = 1, \ldots, M_p \text{ and } j = 1, \ldots, M_s,$$

$$\text{and} \quad \tilde{\boldsymbol{h}}_{i,j}^{s,s} \triangleq \frac{\boldsymbol{h}_{i,j}^{s,s}}{\sqrt{\alpha_j}} \quad \text{for } i \neq j, \; i,j = 1, \ldots, M_s. \qquad (9)$$

Therefore, we can rewrite problem $\tilde{\mathcal{P}}(\boldsymbol{\gamma})$ as follows

$$\tilde{\mathcal{P}}(\boldsymbol{\gamma}) = \begin{cases} \min_{\{\tilde{\boldsymbol{w}}_i^s\}} & \sum_{i=1}^{M_s} \|\tilde{\boldsymbol{w}}_i^s\|^2 \\ \text{s.t.} & \frac{|\tilde{\boldsymbol{h}}_{i,i}^{s,s}\tilde{\boldsymbol{w}}_i^s|^2}{\sum_{j \neq i} |\tilde{\boldsymbol{h}}_{i,j}^{s,s}\tilde{\boldsymbol{w}}_j^s|^2 + \sum_j |\boldsymbol{h}_{i,j}^{s,p}\boldsymbol{w}_j^p|^2 + \sigma_i^s} \geq 1 \quad \text{for } i = 1, \ldots, M_s \\ & \frac{1}{\sum_j |\tilde{\boldsymbol{h}}_{i,j}^{p,s}\tilde{\boldsymbol{w}}_j^s|^2} \geq 1 \quad \text{for } i = 1, \ldots, M_p. \end{cases} \qquad (10)$$

Keeping the interference to the primary users as well as other secondary users low suggests using beamforming vectors with small transmit powers. On the other hand, for small transmit powers the secondary receivers may violate their SINR constraints. Due to this tension, it is not always possible to have feasible solutions for $\{\boldsymbol{w}_i^s\}$. Therefore, we first provide a *necessary* condition for examining the feasibility of $\tilde{\mathcal{P}}(\boldsymbol{\gamma})$. For this purpose, we first define $\tilde{\boldsymbol{w}}_i^p \triangleq \frac{(\boldsymbol{w}_i^p)^H}{\|\boldsymbol{w}_i^p\|^2}$ for $i = 1, \ldots, M_p$, and

$$\boldsymbol{Q} \triangleq \begin{bmatrix} \tilde{\boldsymbol{h}}_{1,1}^{s,s} & \cdots & \tilde{\boldsymbol{h}}_{1,M_s}^{s,s} & \boldsymbol{h}_{1,1}^{s,p} & \cdots & \boldsymbol{h}_{1,M_p}^{s,p} \\ \vdots & \vdots & \vdots & \vdots & \vdots & \vdots \\ \tilde{\boldsymbol{h}}_{M_s,1}^{s,s} & \cdots & \tilde{\boldsymbol{h}}_{M_s,M_s}^{s,s} & \boldsymbol{h}_{M_s,1}^{s,p} & \cdots & \boldsymbol{h}_{M_s,M_p}^{s,p} \\ \tilde{\boldsymbol{h}}_{1,1}^{p,s} & \cdots & \tilde{\boldsymbol{h}}_{1,M_s}^{p,s} & \tilde{\boldsymbol{w}}_1^p & \cdots & 0 \\ \vdots & \vdots & \vdots & \vdots & \vdots & \vdots \\ \tilde{\boldsymbol{h}}_{M_p,1}^{p,s} & \cdots & \tilde{\boldsymbol{h}}_{M_p,M_s}^{p,s} & 0 & \cdots & \tilde{\boldsymbol{w}}_{M_p}^p \end{bmatrix} \quad \text{and} \quad \boldsymbol{T} \triangleq \begin{bmatrix} \tilde{\boldsymbol{w}}_1^s & 0 & \cdots & \cdots & \cdots & 0 \\ 0 & \ddots & 0 & \cdots & \cdots & \vdots \\ \vdots & 0 & \tilde{\boldsymbol{w}}_{M_s}^s & 0 & \cdots & \vdots \\ \vdots & \vdots & 0 & \boldsymbol{w}_1^p & 0 & \vdots \\ \vdots & \vdots & \vdots & 0 & \ddots & 0 \\ 0 & \cdots & \cdots & \cdots & 0 & \boldsymbol{w}_{M_p}^p \end{bmatrix}.$$

Therefore, we can re-write the $M_s + M_p$ constraints of (10) as

$$\min_i \frac{|[\boldsymbol{QT}]_{i,i}|^2}{\sum_{j \neq i} |[\boldsymbol{QT}]_{i,j}|^2 + \boldsymbol{\sigma}(i)} \geq 1, \qquad (11)$$

where $\boldsymbol{\sigma} \triangleq [\sigma_1^s, \ldots, \sigma_{M_s}^s, \mathbf{0}_{1 \times M_p}]$.

**Lemma 1** *Problem $\tilde{\mathcal{P}}(\boldsymbol{\gamma})$ is feasible only if the channel realizations are such that*

$$\mathrm{rank}(\boldsymbol{Q}) \geq \frac{M_s + M_p}{2}.$$

*Proof:* See Appendix B. ∎

A simple sufficient condition for feasibility is given next. It uses the fact that if each secondary transmitter can employ a beamformer which causes zero interference to all unintended receivers but yields a positive signal strength at the intended receiver, then any given set of SINRs are achievable. Let $\boldsymbol{H}_{i,\bar{i}}^s, \forall\, i$ be the $(M_s + M_p - 1) \times N_s$ matrix whose rows are the outgoing channel vectors of the $i^{th}$ secondary transmitter corresponding to all its unintended receivers.

**Lemma 2** *Problem $\tilde{\mathcal{P}}(\boldsymbol{\gamma})$ is feasible for any finite $\boldsymbol{\gamma}$ if the channel realizations are such that for each secondary transmitter $i$*

$$(\boldsymbol{h}_{i,i}^{s,s})^H \notin \mathrm{Range}\{(\boldsymbol{H}_{i,\bar{i}}^s)^H\}.$$

In the subsequent analysis we assume that the optimization problem $\tilde{\mathcal{P}}(\boldsymbol{\gamma})$ is strictly feasible. Note that the optimization problem $\tilde{\mathcal{P}}(\boldsymbol{\gamma})$ is not convex in its direct form. Nevertheless, we can show that strong duality holds for this problem by arguing that this problem can be transformed to an SOCP which has a linear objective function and second order cone constraints and is convex. The same idea has been employed for precoder design in downlink MISO point-to-multi-point channels [6].

**Lemma 3** *Strong duality holds for $\tilde{\mathcal{P}}(\boldsymbol{\gamma})$, i.e., it has zero duality gap with its Lagrangian dual.*

*Proof:* See Appendix C. ∎

When all the channels are known to a central agent, as shown in the Appendix, the problem $\tilde{\mathcal{P}}(\boldsymbol{\gamma})$ can be solved efficiently as an SOCP. In a distributed network, however, each secondary transmitter has access only to limited channel state information. Here, we do not assume global channel state information at each secondary user. Instead we assume that primary users have no side-information about the secondary channel states and each secondary user only knows its outgoing (forward) channels to all the secondary and the primary receivers. In addition, each secondary user also knows all the incoming channel vectors seen by its intended receiver from other secondary users as well as the effective noise variance (i.e., thermal noise variance plus the interference due to primary transmissions) at its intended receiver. Lack of complete knowledge of channel state information at the secondary transmitters necessitates developing distributed algorithms to be run by the secondary users in a decentralized fashion with limited message passing among themselves. In the sequel we develop a distributed scheme which yields the global optimal solution of $\tilde{\mathcal{P}}(\boldsymbol{\gamma})$ by solving its Lagrangian dual problem.

We construct the *partial* Lagrangian function of the problem $\tilde{\mathcal{P}}(\boldsymbol{\gamma})$ given in (10) by dualizing only the interference margins. For a non-negative set of multipliers $\boldsymbol{\lambda} \triangleq [\lambda_1 \ldots \lambda_{M_p}]^T$, associated with the interference margins, the Lagrangian function is given by

$$\begin{aligned} L(\{\tilde{\boldsymbol{w}}_i^s\}, \boldsymbol{\lambda}) &= \sum_{i=1}^{M_s} \|\tilde{\boldsymbol{w}}_i^s\|^2 + \sum_{j=1}^{M_p} \lambda_j \left[ \sum_{i=1}^{M_s} |\tilde{\boldsymbol{h}}_{j,i}^{p,s} \tilde{\boldsymbol{w}}_i^s|^2 - 1 \right] \\ &= \sum_{i=1}^{M_s} (\tilde{\boldsymbol{w}}_i^s)^H \left[ \boldsymbol{I} + \sum_{j=1}^{M_p} (\tilde{\boldsymbol{h}}_{j,i}^{p,s})^H \tilde{\boldsymbol{h}}_{j,i}^{p,s} \lambda_j \right] \tilde{\boldsymbol{w}}_i^s - \sum_{j=1}^{M_p} \lambda_j. \end{aligned}$$

It is easy to verify that $\boldsymbol{I} + \sum_{j=1}^{M_p} (\tilde{\boldsymbol{h}}_{j,i}^{p,s})^H \tilde{\boldsymbol{h}}_{j,i}^{p,s} \lambda_j$ is positive definite for all $i$, and by the Cholesky decomposition we know that there exists an invertible triangular matrix $\boldsymbol{U}_i$ such that

$$\boldsymbol{U}_i^H \boldsymbol{U}_i = \boldsymbol{I} + \sum_{j=1}^{M_p} (\tilde{\boldsymbol{h}}_{j,i}^{p,s})^H \tilde{\boldsymbol{h}}_{j,i}^{p,s} \lambda_j, \qquad (12)$$

where $\boldsymbol{U}_i$, although not explicitly shown, depends on $\boldsymbol{\lambda}$. Therefore, since we have incorporated only the interference constraints in formulating the Lagrangian function, the Lagrange dual function is given by

$$g(\boldsymbol{\lambda}) = \begin{cases} \min_{\{\tilde{\boldsymbol{w}}_i^s\}} & \sum_{i=1}^{M_s} (\tilde{\boldsymbol{w}}_i^s)^H \boldsymbol{U}_i^H \boldsymbol{U}_i \tilde{\boldsymbol{w}}_i^s - \sum_{j=1}^{M_p} \lambda_j \\ \text{s.t.} & \sum_{j=1}^{M_s} |\tilde{\boldsymbol{h}}_{i,j}^{s,s} \tilde{\boldsymbol{w}}_j^s|^2 - 2|\tilde{\boldsymbol{h}}_{i,i}^{s,s} \tilde{\boldsymbol{w}}_i^s|^2 + a_i^s \leq 0 \quad \text{for } i = 1, \ldots, M_s, \end{cases} \qquad (13)$$

where we have defined

$$a_i^s \triangleq \sum_j |\boldsymbol{h}_{i,j}^{s,p} \boldsymbol{w}_j^p|^2 + \sigma_i^s \quad \text{for } i = 1, \ldots, M_s. \qquad (14)$$

By further defining $\hat{\boldsymbol{w}}_i^s \triangleq \boldsymbol{U}_i \tilde{\boldsymbol{w}}_i^s$ and $\hat{\boldsymbol{h}}_{j,i}^{s,s} \triangleq \tilde{\boldsymbol{h}}_{j,i}^{s,s} \boldsymbol{U}_i^{-1}$ we get

$$g(\boldsymbol{\lambda}) = \begin{cases} \min_{\{\hat{\boldsymbol{w}}_i^s\}} & \sum_{i=1}^{M_s} (\hat{\boldsymbol{w}}_i^s)^H \hat{\boldsymbol{w}}_i^s - \sum_{j=1}^{M_p} \lambda_j \\ \text{s.t.} & \dfrac{|\hat{\boldsymbol{h}}_{i,i}^{s,s} \hat{\boldsymbol{w}}_i^s|^2}{\sum_{j \neq i} |\hat{\boldsymbol{h}}_{i,j}^{s,s} \hat{\boldsymbol{w}}_j^s|^2 + a_i^s} \geq 1 \quad \text{for } i = 1, \ldots, M_s. \end{cases} \qquad (15)$$

Note that for any given set of $\{\lambda_i\}$, the term $\sum_{j=1}^{M_p} \lambda_j$ is constant and the problem in (15) is equivalent to minimizing $\sum_{i=1}^{M_s} (\hat{\boldsymbol{w}}_i^s)^H \hat{\boldsymbol{w}}_i^s$. It can now be verified that the problem in (15) is equivalent to the optimization problem in [7] which considers beamforming design over downlink MISO multi-cell networks. In particular, [7] aims to minimize the weighted sum power subject to SINR constraints for the users in different cells and proposes an algorithm for solving this problem in a distributed way (the base stations do not collaborate for designing their beamformers). The underlying idea of this distributed algorithm is to exploit the uplink-downlink duality in multi-antenna transmissions. By leveraging the approach in [7], we can show that for any given set of $\{\lambda_i\}$, $g(\boldsymbol{\lambda})$ can be computed in a distributed way.

On the other hand, by using the result of Lemma 3 that strong duality holds for $\tilde{\mathcal{P}}(\boldsymbol{\gamma})$, we know that

$$\max_{\boldsymbol{\lambda} \succeq \boldsymbol{0}} g(\boldsymbol{\lambda}) = \tilde{\mathcal{P}}(\boldsymbol{\gamma}). \tag{16}$$

Therefore, we utilize the *subgradient method* which provides a simple algorithm for minimizing such convex problems with possibly non-differentiable objective function [12]. We apply the subgradient method on $\hat{g}(\boldsymbol{\lambda}) \triangleq -g(\boldsymbol{\lambda})$ (which is convex). According to the subgradient method, we use the following iterative procedure to minimize $\hat{g} : \mathbb{R}^{M_p} \to \mathbb{R}$.

$$\boldsymbol{\lambda}^{(k+1)} = \boldsymbol{\lambda}^{(k)} - \mu_k \boldsymbol{s}^{(k)},$$

where $\boldsymbol{s}^{(k)}$ is *any* subgradient of $\hat{g}$ at $\boldsymbol{\lambda}^{(k)}$. Recall that a subgradient of $\hat{g}$ at $\boldsymbol{\lambda}^{(k)}$ is any vector $\boldsymbol{s}^{(k)}$ that satisfies $\hat{g}(\boldsymbol{\lambda}') \geq \hat{g}(\boldsymbol{\lambda}^{(k)}) + (\boldsymbol{s}^{(k)})^T(\boldsymbol{\lambda}' - \boldsymbol{\lambda}^{(k)})$, $\forall \boldsymbol{\lambda}'$ and $\mu_k > 0$ is the $k^{th}$ step size. Since the subgradient method is not a descent method, at each iteration we need to keep track of the best point found thus far, i.e.,

$$\hat{g}_{\text{best}}^{(k)} = \min\{\hat{g}_{\text{best}}^{(k-1)}, \hat{g}(\boldsymbol{\lambda}^{(k)})\}.$$

The subgradient method is guaranteed to converge to the optimal value for *non-summable diminishing* step sizes, i.e., $\lim_{k \to \infty} \mu_k = 0$ and $\sum_{k=1}^{\infty} \mu_k = \infty$ [12]. Hence, to be able to use the subgradient method we need to find a valid subgradient and a step size. We select $\mu_k = \frac{1}{k}$ which can be easily seen to be a non-summable diminishing step size. Also we argue that for any point $\boldsymbol{\lambda}^{(k)}$, $\boldsymbol{s}^{(k)} \triangleq [s_1^{(k)}, \ldots, s_{M_p}^{(k)}]$ such that

$$s_j^{(k)} = 1 - \sum_{i=1}^{M_s} |\tilde{\boldsymbol{h}}_{j,i}^{p,s} \boldsymbol{\omega}_i|^2, \quad \text{for } j = 1, \ldots, M_p, \quad \text{where} \quad \{\boldsymbol{\omega}_i\} = \arg\min_{\{\tilde{\boldsymbol{w}}_i^s\} \in \mathcal{D}_{\{\tilde{\boldsymbol{h}}_{i,j}^{s,s}\}}} L(\{\tilde{\boldsymbol{w}}_i^s\}, \boldsymbol{\lambda}^{(k)}), \tag{17}$$

is a valid subgradient, since $\forall \boldsymbol{\lambda}'$

$$
\begin{aligned}
\hat{g}(\boldsymbol{\lambda}') &= -g(\boldsymbol{\lambda}') = - \min_{\{\tilde{\boldsymbol{w}}_i^s\} \in \mathcal{D}_{\{\tilde{\boldsymbol{h}}_{i,j}^{s,s}\}}} L(\{\tilde{\boldsymbol{w}}_i^s\}, \boldsymbol{\lambda}') \\
&\geq -L(\{\boldsymbol{\omega}_i\}, \boldsymbol{\lambda}') = -\left\{ \sum_{i=1}^{M_s} \|\boldsymbol{\omega}_i\|^2 + \sum_{j=1}^{M_p} \lambda'_j \left[ \sum_{i=1}^{M_s} |\tilde{\boldsymbol{h}}_{j,i}^{p,s} \boldsymbol{\omega}_i|^2 - 1 \right] \right\} \\
&= -\left\{ \sum_{i=1}^{M_s} \|\boldsymbol{\omega}_i\|^2 + \sum_{j=1}^{M_p} \lambda_j^{(k)} \left[ \sum_{i=1}^{M_s} |\tilde{\boldsymbol{h}}_{j,i}^{p,s} \boldsymbol{\omega}_i|^2 - 1 \right] \right\} + \sum_{j=1}^{M_p} (\lambda'_j - \lambda_j^{(k)}) \left[ 1 - \sum_{i=1}^{M_s} |\tilde{\boldsymbol{h}}_{j,i}^{p,s} \boldsymbol{\omega}_i|^2 \right] \\
&= - \min_{\{\tilde{\boldsymbol{w}}_i^s\} \in \mathcal{R}_{\{\boldsymbol{h}_{i,j}^{s,s}\}}} L(\{\tilde{\boldsymbol{w}}_i^s\}, \boldsymbol{\lambda}^{(k)}) + (\boldsymbol{s}^{(k)})^T (\boldsymbol{\lambda}' - \boldsymbol{\lambda}^{(k)}) = \hat{g}(\boldsymbol{\lambda}^{(k)}) + (\boldsymbol{s}^{(k)})^T (\boldsymbol{\lambda}' - \boldsymbol{\lambda}^{(k)}) \quad (18)
\end{aligned}
$$

Therefore, by the aforementioned choices for $\boldsymbol{s}^{(k)}$ and $\mu_k$ the subgradient method is guaranteed to converge to the optimal value and thus, yields the minimum of $-g(\boldsymbol{\lambda})$ or the maximum of $g(\boldsymbol{\lambda})$ which is the optimal value of $\tilde{\mathcal{P}}(\boldsymbol{\gamma})$. Algorithm 1 summarizes the steps involved in obtaining $\tilde{\mathcal{P}}(\boldsymbol{\gamma})$.

In practice, for implementing Algorithm 1 we allow a specified maximum number of iterations. If no convergence is observed, we declare the power minimization problem to be infeasible and set $\mathcal{P}(\boldsymbol{\gamma}) = \infty$.

Finally, in order to solve the optimization problem $\mathcal{R}(P_0)$ we exploit the relationship between $\mathcal{P}(\boldsymbol{\rho})$ and $\mathcal{R}(P_0)$ given in Theorem 1. Using Algorithm 1, along with a bi-section search, in Algorithm 2 we provide the steps for solving $\mathcal{R}(P_0)$. For initializing the algorithm we need lower and upper bounds on the (optimal) $\mathcal{R}(P_0)$, which we denote by $\rho_{\min}$ and $\rho_{\max}$, respectively. For both bounds, we use beamforming vectors obtained via channel matching, i.e., we set $\boldsymbol{w}_i^s$ to be a scalar multiple of $(\boldsymbol{h}_{i,i}^{s,s})^H / \|\boldsymbol{h}_{i,i}^{s,s}\|$. In particular, for obtaining $\rho_{\min}$, we set $\boldsymbol{w}_i^s = \sqrt{\hat{\alpha}} (\boldsymbol{h}_{i,i}^{s,s})^H / \|\boldsymbol{h}_{i,i}^{s,s}\|, \ \forall \ i$, where $\hat{\alpha}$ is the largest positive scalar such that the power and margin constraints are satisfied. For obtaining $\rho_{\max}$, we assume the presence of a genie which ensures that the transmission intended for any particular secondary receiver causes no interference to any other receiver and can use all the available power, so that the optimal secondary beamformers are $\{\frac{P_0 (\boldsymbol{h}_{i,i}^{s,s})^H}{\alpha_i \|\boldsymbol{h}_{i,i}^{s,s}\|}\}$. Note that Algorithm 2 always returns a feasible $\rho$ and $\{\boldsymbol{w}_i^s\}$.

## 4.2 Maximum Likelihood Decoders

In this setup each secondary receiver deploys the MLD which jointly decodes all transmitted secondary codewords and is optimal only with respect to the joint error probability. Recall that each secondary receiver is interested in recovering only the codeword transmitted by its designated secondary transmitter. Therefore, the $i^{th}$ receiver secondary will declare an error if and only if it fails to decode the codeword of its desired secondary user $i$.

For each user $i$ we define the sets $\mathcal{V}_i^k$, for $k = 1, \ldots, 2^{M_s-1}$, to be all possible subsets of $\{1, \ldots, M_s\}$ which contain $i$, i.e., $i \in \mathcal{V}_i^k$, $\forall k$. In the following lemma we provide a sufficient condition for a rate vector $\boldsymbol{R}$ to be decodable. This condition explicitly incorporates the effect of the channel coefficients and beamforming vectors on the decodability of rate vector $\boldsymbol{R}$.

**Lemma 4** *When all secondary receivers employ MLD, a sufficient condition for a rate vector $\boldsymbol{R}$ to be decodable is*

$$\forall k, \quad \log\left(1 + \frac{\sum_{j \in \mathcal{V}_i^k} |\boldsymbol{h}_{i,j}^{s,s} \boldsymbol{w}_j^s|^2}{\sum_j |\boldsymbol{h}_{i,j}^{s,p} \boldsymbol{w}_j^p|^2 + \sigma_i^s}\right) \geq \sum_{j \in \mathcal{V}_i^k} R_j \quad for \ i = 1, \ldots, M_s. \tag{19}$$

*Proof:* See Appendix D. ∎

Note that the usual achievable rate region for the MLD is obtained under the restriction that all codewords can be reliably decoded. In particular, this region is the set of all rate vectors, $\boldsymbol{R}$ which satisfy

$$\log\left(1 + \frac{\sum_{j \in \mathcal{S}} |\boldsymbol{h}_{i,j}^{s,s} \boldsymbol{w}_j^s|^2}{\sum_j |\boldsymbol{h}_{i,j}^{s,p} \boldsymbol{w}_j^p|^2 + \sigma_i^s}\right) \geq \sum_{j \in \mathcal{S}} R_j, \quad \forall \ i, \ \mathcal{S} \subseteq \{1, \cdots, M_s\} \ \& \ \mathcal{S} \neq \phi. \tag{20}$$

The condition we derived in (19) is more relaxed compared to the one in (20), since we only need the $i^{th}$ user to be decodable at the $i^{th}$ receiver. While we do not prove that the condition in (19) is also necessary, we note that using the approach in [13], it can be argued that a decoding error for user $i$ at receiver $i$ is very likely if the condition in (19) is not satisfied.

Next, we define the vector $\boldsymbol{w} \triangleq [(\boldsymbol{w}_1^s)^T, \ldots, (\boldsymbol{w}_{M_s}^s)^T]^T$. Also, corresponding to each subset $\mathcal{V}_i^k$ we construct an $N_s M_s \times N_s M_s$ square matrix $\boldsymbol{B}_i^k$ that consists of $M_s^2$ square sub-matrices each of dimension $N_s \times N_s$. The $(j,j)^{th}$ sub-matrix in $\boldsymbol{B}_i^k$ for each $j \in \mathcal{V}_i^k$ is $-(\boldsymbol{h}_{i,j}^{s,s})^H \boldsymbol{h}_{i,j}^{s,s}$ and other sub-matrices are $\boldsymbol{0}$. Further, for a rate vector $\boldsymbol{\rho}$, corresponding to each $\mathcal{V}_i^k$ we define the scalar

$$c_i^k = \left(1 - 2^{\sum_{j \in \mathcal{V}_i^k} \rho_j}\right)\left(\sum_j |\boldsymbol{h}_{i,j}^{s,p} \boldsymbol{w}_j^p|^2 + \sigma_i^s\right).$$

Therefore, the conditions in (19) can be rewritten as $\boldsymbol{w}^H \boldsymbol{B}_i^k \boldsymbol{w} - c_i^k \leq 0, \forall i, k$. The primary interference margin constraints $\boldsymbol{J} \preceq \boldsymbol{\beta}$ can also be restated as $\boldsymbol{w}^H \tilde{\boldsymbol{B}}_i \boldsymbol{w} - \beta_i \leq 0$, where $\tilde{\boldsymbol{B}}_i$ consists of $M_s^2$ square sub-matrices each of dimension $N_s \times N_s$, such that the $(j,j)^{th}$ sub-matrix, for $1 \leq j \leq M_s$, is $(\boldsymbol{h}_{i,j}^{p,s})^H \boldsymbol{h}_{i,j}^{p,s}$ and other sub-matrices are $\boldsymbol{0}$.

By defining $\boldsymbol{A} \triangleq \mathrm{diag}[\alpha_1, \ldots, \alpha_{M_s}] \otimes \boldsymbol{I}_{N_s}$ the power minimization problem is given by

$$\begin{cases} \min_{\boldsymbol{w}} & \boldsymbol{w}^H \boldsymbol{A} \boldsymbol{w}, \\ \text{s.t.} & \boldsymbol{w}^H \boldsymbol{B}_i^k \boldsymbol{w} - c_i^k \leq 0 \quad \forall i, k, \\ & \boldsymbol{w}^H \tilde{\boldsymbol{B}}_i \boldsymbol{w} - \beta_i \leq 0 \quad \forall i. \end{cases} \quad (21)$$

The problem in (21) is a non-convex problem with a quadratic objective and *indefinite* quadratic constraints which in its general form is an NP hard problem [8]. Several approaches have been developed for solving this problem sub-optimally, including a convex relaxation approach which converts the problem to a semi-definite program (SDP) described as follows.

By defining $\boldsymbol{X} \triangleq \boldsymbol{w}\boldsymbol{w}^H$, we obtain the problem

$$\begin{cases} \min_{\boldsymbol{X}} & \mathrm{tr}(\boldsymbol{X}\boldsymbol{A}), \\ \text{s.t.} & \mathrm{tr}(\boldsymbol{B}_i^k \boldsymbol{X}) - c_i^k \leq 0 \quad \forall i, k, \\ & \mathrm{tr}(\tilde{\boldsymbol{B}}_i \boldsymbol{X}) - \beta_i \leq 0 \quad \forall i, \\ & \boldsymbol{X} = \boldsymbol{w}\boldsymbol{w}^H. \end{cases} \quad (22)$$

By relaxing the constraint $\boldsymbol{X} = \boldsymbol{w}\boldsymbol{w}^H$ and replacing it with the convex constraint $\boldsymbol{X} \succeq \boldsymbol{0}$ the problem becomes an SDP which is a convex problem and can be solved efficiently. We denote the solution of the SDP by $\boldsymbol{X}^*$. Note that a feasible solution for (22) should be a rank-1 matrix, which is not necessarily true for the solution of the SDP. There are several methods to recover a rank-1 solution, see [14] where such methods are discussed in the context of multi-cast beamforming. Here we adopt a simple approach and first determine the dominant eigenvector of $\boldsymbol{X}$ and denote it by $\tilde{\boldsymbol{w}} = [(\tilde{\boldsymbol{w}}_1^s)^T, \ldots, (\tilde{\boldsymbol{w}}_{M_s}^s)^T]^T$. Then for finding a feasible $\boldsymbol{w}$ we set $\boldsymbol{w}_i^s = \sqrt{p_i} \tilde{\boldsymbol{w}}_i^s / \|\tilde{\boldsymbol{w}}_i^s\|$, where $1 \leq i \leq M_s$ and $\{p_i\}$ are positive scaling factors which can be found by solving a linear program (LP) which minimizes the weighted sum power under the constraints that the rate and margin constraints are satisfied. In case either the SDP or the LP are infeasible, we declare the power minimization problem to be infeasible. Finally, in order to solve the rate optimization problem we can exploit the relationship between $\mathcal{P}(\boldsymbol{\rho})$ and $\mathcal{R}(P_0)$ given in Theorem 1 and use the aforementioned power minimization technique along with a bi-section search similar to the one provided for the MMSE receiver in the previous section. While such a method provides a sub-optimal design of the beamforming vectors for the MLD, a drawback is that it cannot be implemented in a distributed way.

## 4.3 Unconstrained Group Decoders

In this section we assume that each secondary receiver is equipped with the unconstrained group decoder (UGD) [9]. Note that a drawback of the MLD is that it decodes all secondary users which entails a high decoding complexity and can degrade performance in cases when some of the other users are best treated as noise. In fact a rate vector decodable using the MMSE decoder at each receiver need not necessarily be decodable upon using the MLD at each receiver. The UGD on the other hand, may decode the designated user jointly with any arbitrary subset of other secondary users, after suppressing or canceling any other subsets. As a result, any rate vector which is decodable by MMSE or ML decoders, it is also decodable by the UGD. In the following remark we first state a sufficient condition for a subset of users to be decodable when the UGD is deployed at each secondary receiver.

**Remark 1** *A subset of users $\mathcal{U} \subseteq \{1, \ldots, M_s\}$ is decodable at the $i^{th}$ receiver under the rate assignment $\boldsymbol{R}$ if for all non-empty sets $\mathcal{A} \subseteq \mathcal{U}$ we have*

$$\log\left(1 + \frac{\sum_{j \in \mathcal{A}} |\boldsymbol{h}_{i,j}^{s,s} \boldsymbol{w}_j^s|^2}{\sum_{j \notin \mathcal{U}} |\boldsymbol{h}_{i,j}^{s,s} \boldsymbol{w}_j^s|^2 + \sum_{j=1}^{M_p} |\boldsymbol{h}_{i,j}^{s,p} \boldsymbol{w}_j^p|^2 + \sigma_i^s}\right) \geq \sum_{j \in \mathcal{A}} R_j. \tag{23}$$

*Moreover, if for an ordered partition $\{\mathcal{G}_1, \cdots, \mathcal{G}_p\}$ of any subset $\mathcal{U} \subseteq \{1, \ldots, M_s\}$, the set $\mathcal{G}_j$, where $1 \leq j \leq p$, is decodable at the $i^{th}$ receiver after expurgating users in $\{\mathcal{G}_1, \cdots, \mathcal{G}_{j-1}\}$, i.e.,*

$$\log\left(1 + \frac{\sum_{j \in \mathcal{A}} |\boldsymbol{h}_{i,j}^{s,s} \boldsymbol{w}_j^s|^2}{\sum_{j \notin \cup_{q=1}^{j} \mathcal{G}_q} |\boldsymbol{h}_{i,j}^{s,s} \boldsymbol{w}_j^s|^2 + \sum_{j=1}^{M_p} |\boldsymbol{h}_{i,j}^{s,p} \boldsymbol{w}_j^p|^2 + \sigma_i^s}\right) \geq \sum_{j \in \mathcal{A}} R_j, \forall \mathcal{A} \subseteq \mathcal{G}_j, 1 \leq j \leq p, \tag{24}$$

*then we have that the set $\mathcal{U} = \cup_{q=1}^{p} \mathcal{G}_q$ is also decodable.*

As a result, a sufficient condition for user $i$ to be decodable at the $i^{th}$ receiver when the UGD is deployed is that *there exists a decodable subset $\mathcal{U}$ containing $i$*. It is informative to notice here that the worst case decoding complexity of the UGD is equal to the decoding complexity of the MLD. Also, suppose no subset containing $i$ satisfies the condition in (23). In this case, using the results from [9] we can conclude that user $i$ belongs to a unique undecodable set at receiver $i$, so that a decoding error for user $i$ is very likely if we attempt to decode $i$ using the UGD.

For the purpose of making the optimization formulation more intuitive we assume identical rate weighting for all secondary users, i.e., for all users we set $\rho = \rho_i$. As a result, the optimal design assigns identical rates to all users so that $R_j = R, \forall j$. Generalization to an arbitrary weight vector $\boldsymbol{\rho}$ is possible by following the same approach. The following remark is helpful in identifying a decodable subset (if any) at receiver $i$ containing user $i$.

**Remark 2** If the $i^{th}$ secondary user is decodable at receiver $i$, then all users $k \in \{1, \ldots, M_s\}$ such that $|h_{i,k}^{s,s} w_k^s| \geq |h_{i,i}^{s,s} w_i^s|$ are also decodable at the $i^{th}$ receiver.

The remark above conveys that for finding a decodable subset containing $i$, we need to only consider those subsets that contain $i$ as well as all $k$ such that $|h_{i,k}^{s,s} w_k^s| \geq |h_{i,i}^{s,s} w_i^s|$. Let $\pi_i(\cdot)$ be a permutation operator on the indices of the users such that

$$|h_{i,\pi_i(1)}^{s,s} w_{\pi_i(1)}^s| \geq \cdots \geq |h_{i,\pi_i(M_s)}^{s,s} w_{\pi_i(M_s)}^s|. \tag{25}$$

By using Remarks 1 and 2 and some simple manipulations, we can compactly express the sufficient condition for user $i$ to be decodable as follows.

**Remark 3** User $i$ is decodable if $\max_{\pi_i^{-1}(i) \leq p \leq M_s} f(i,p) \geq R$, where

$$f(i,p) \triangleq \min_{1 \leq q \leq p} \left\{ \frac{1}{p-q+1} \log \left( 1 + \frac{\sum_{m=q}^{p} |h_{i,\pi_i(m)}^{s,s} w_{\pi_i(m)}^s|^2}{\sum_{m=p+1}^{M_s} |h_{i,\pi_i(m)}^{s,s} w_{\pi_i(m)}^s|^2 + \sum_j |h_{i,j}^{s,p} w_j^p|^2 + \sigma_i^s} \right) \right\} \quad \forall i. \tag{26}$$

Also, according to the above remark the optimal decoding set containing the $i^{th}$ user that supports the largest worst-case rate is the set $\{\pi_i(1), \ldots, \pi_i(p^*)\}$ where $p^* = \arg\max_{\pi_i^{-1}(i) \leq p \leq M_s} f(i,p)$. Then, the rate optimization problem is given by

$$\begin{cases} \max_{\{w_i^s\}} & R, \\ \text{s.t.} & \sum_{i=1}^{M_s} \alpha_i \|w_i^s\|^2 \leq P_0, \\ & \max_{\pi_i^{-1}(i) \leq p \leq M_s} f(i,p) \geq R \quad \forall i, \\ & J_i \leq \beta_i, \quad i = 1, \ldots, M_p. \end{cases}$$

This beamforming optimization problem for UGDs is a non-linear non-convex problem for which an optimal solution cannot be guaranteed even in a centralized setup. Note that in the case that $\rho_i$ are not identical, the problem is even more involved.

## 5 Distributed Weighted Max-Min Fair Rate Allocation

As shown in Sections 4.2 and 4.3, solving the beamforming and rate optimization problems for the MLD and UGD, respectively, are non-linear non-convex problems for which even centralized algorithms are not guaranteed to yield globally optimal solutions. Motivated by this fact and more importantly by the necessity for having a distributed algorithm, we propose an alternative two-stage suboptimal approach. We first obtain the beamforming vectors via Algorithms 1 and 2 which provide the optimal beamformers for the MMSE receivers. In the second stage, for the given choice of beamformers, we exploit the fact that the MLD or UGD is used at each secondary receiver and allocate excess rates to secondary users in a distributed fashion.

We denote the optimal beamforming vectors yielded by Algorithm 2 by $\{\boldsymbol{w}_i^*\}_{i=1}^{M_s}$. We use $h_{i,j} \in \mathbb{C}$ to denote the combined effect of beamforming vectors $(\boldsymbol{w}_j^*)$ and channel coefficients $(\boldsymbol{h}_{i,j}^{s,s})$. Thus,

$$h_{i,j} = \boldsymbol{h}_{i,j}^{s,s} \boldsymbol{w}_j^* \quad \text{for } i,j = 1,\ldots,M_s.$$

Further, in all rate allocation algorithms proposed in the sequel, we will guarantee that each user receives at-least a minimum rate and the vector of minimum rates is denoted by $\boldsymbol{R}^{\min}$. Such minimum rate vector for instance, can be the vector of rates achieved by using MMSE decoders.

We define $\boldsymbol{x} \triangleq [x_1^s, \ldots, x_{M_s}^s]^T$ as the vector of information symbols transmitted by all secondary transmitters and also define $\boldsymbol{h}^i \triangleq [h_{i,1}, \ldots, h_{i,M_s}]$. Therefore, the signal received by the $i^{th}$ secondary receiver is $y_i = \boldsymbol{h}^i \boldsymbol{x} + z_i$, where $z_i \in \mathbb{C}$ accounts for the Gaussian noise as well as the interference seen from the primary users. Without loss of generality, we assume $z_i \sim \mathcal{CN}(0,1)$. As before we assume that $x_i^s$, the information symbol of the $i^{th}$ user, has unit power and is drawn from a Gaussian alphabet.

We use $\mathcal{K} = \{1, \ldots, M_s\}$ to refer to the set of all secondary users and construct the vector $\boldsymbol{h}_{\mathcal{A}}^i \triangleq [h_{i,j}]_{j \in \mathcal{A}}$ having the scalars $h_{i,j}$, where $j \in \mathcal{A}$, as its elements. $\boldsymbol{R}_{\mathcal{A}}$ denotes the rate vector of the users with indices in $\mathcal{A} \subseteq \mathcal{K}$. For any two disjoint subsets $\mathcal{A}$ and $\mathcal{B}$ of $\mathcal{K}$, let $\mathcal{C}(\boldsymbol{h}^i, \mathcal{A}, \mathcal{B})$ denote an instantaneous achievable rate region for the users in $\mathcal{A}$ which are jointly decoded while treating (suppressing) the users in $\mathcal{B}$ as Gaussian interferers. $\mathcal{C}(\boldsymbol{h}^i, \mathcal{A}, \mathcal{B})$ is given by

$$\mathcal{C}(\boldsymbol{h}^i, \mathcal{A}, \mathcal{B}) = \left\{ \boldsymbol{r} \in \mathbb{R}_+^{|\mathcal{A}|} \;\Big|\; \sum_{j \in \mathcal{D}} r_j \leq \log\det\left[\boldsymbol{I} + \boldsymbol{h}_{\mathcal{D}}^{i\,H}\left(1 + \boldsymbol{h}_{\mathcal{B}}^i \boldsymbol{h}_{\mathcal{B}}^{i\,H}\right)^{-1} \boldsymbol{h}_{\mathcal{D}}^i\right] \;\forall \mathcal{D} \subseteq \mathcal{A} \right\}. \tag{27}$$

Let also $\{\mathcal{G}_i, \mathcal{G}_i^c\}$ denote a partition of $\mathcal{K}$ such that the users in $\mathcal{G}_i$ are jointly decoded after treating those in $\mathcal{G}_i^c$ as noise. Therefore, a rate vector $\boldsymbol{R}$ is decodable using the UGD if there exist sets $\{\mathcal{G}_1, \ldots, \mathcal{G}_{M_s}\}$ such that $i \in \mathcal{G}_i$, $\forall\, i$ and

$$\boldsymbol{R}_{\mathcal{G}_i} \in \mathcal{C}(\boldsymbol{h}^i, \mathcal{G}_i, \mathcal{G}_i^c), \qquad \text{for } i = 1, \ldots, M_s. \tag{28}$$

It is informative to compare the decodability condition for the UGD derived in (28) with that for the ML decoder in (19). Note that if we take $\mathcal{G}_i = \{1, \cdots, M_s\}$ in (28), we obtain the condition in (20) rather than the one in (19). In fact for any two disjoint subsets $\mathcal{A}$ and $\mathcal{B}$ of $\mathcal{K}$ such that

$i \in \mathcal{A}$, by defining

$$\tilde{\mathcal{C}}(\boldsymbol{h}^i, \mathcal{A}, \mathcal{B}) = \left\{ \boldsymbol{r} \in \mathbb{R}_+^{|\mathcal{A}|} \;\middle|\; \sum_{j \in \mathcal{D}} r_j \leq \log \det \left[ \boldsymbol{I} + \boldsymbol{h}_{\mathcal{D}}^{i\,H} \left(1 + \boldsymbol{h}_{\mathcal{B}}^i \boldsymbol{h}_{\mathcal{B}}^{i\,H}\right)^{-1} \boldsymbol{h}_{\mathcal{D}}^i \right] \forall \mathcal{D} \subseteq \mathcal{A} : i \in \mathcal{D} \right\}, \quad (29)$$

we could instead say that $\boldsymbol{R}$ is decodable (using the UGD at each secondary receiver) if there exist sets $\{\mathcal{G}_1', \ldots, \mathcal{G}_{M_s}'\}$ such that

$$i \in \mathcal{G}_i' \;\&\; \boldsymbol{R}_{\mathcal{G}_i'} \in \tilde{\mathcal{C}}(\boldsymbol{h}^i, \mathcal{G}_i', \mathcal{G}_i'^c), \qquad \text{for } i = 1, \ldots, M_s. \quad (30)$$

However, using the fact that the UGD allows us to jointly decode any subset $\mathcal{G}_i$ of $\mathcal{K}$ such that $i \in \mathcal{G}_i$, it can be shown that the conditions in (28) and (30) are identical and hence (28) is more relaxed compared to the one derived for the ML decoder in (19).

## 5.1 Unconstrained Group Decoders

We consider increasing the rates of all users based on some pre-determined priority, given any decodable rate vector $\boldsymbol{R}^{\min} \triangleq [R_1^{\min}, \ldots, R_{M_s}^{\min}]$. We assign different priorities to the secondary users through the factors $\boldsymbol{\rho} \triangleq [\rho_1, \ldots, \rho_{M_s}]$. By denoting the excess rate to be assigned to the $i^{th}$ user by $r_i$, our objective is to maximize $\min_i r_i/\rho_i$ such that $\boldsymbol{R}^{\min} + \boldsymbol{r}$ remains decodable where $\boldsymbol{r} \triangleq [r_1, \ldots, r_{M_s}]$. For any secondary receiver $i$ and any two non-empty disjoint subsets $\mathcal{A}, \mathcal{B}$ of $\mathcal{K}$ we define

$$\theta(\boldsymbol{h}^i, \mathcal{A}, \mathcal{B}, \boldsymbol{R}^{\min}, \boldsymbol{\rho}) \triangleq \begin{cases} \max & \min_k \frac{r_k}{\rho_k} \\ \text{s.t.} & \boldsymbol{r}_{\mathcal{A}} + \boldsymbol{R}_{\mathcal{A}}^{\min} \in \mathcal{C}(\boldsymbol{h}^i, \mathcal{A}, \mathcal{B}) \end{cases}, \quad (31)$$

which picks a rate-vector within the rate region $\mathcal{C}(\boldsymbol{h}^i, \mathcal{A}, \mathcal{B})$ and achieves weighted max-min fairness for the users in $\mathcal{A}$. To solve (31), we note that the region $\mathcal{C}(\boldsymbol{h}^i, \mathcal{A}, \mathcal{B})$ can be shown to be a polymatroid [15, Lemma 3.4] with the rank function

$$f(\mathcal{S}) \triangleq \log \det \left[ \boldsymbol{I} + \boldsymbol{h}_{\mathcal{S}}^{i\,H} \left(1 + \boldsymbol{h}_{\mathcal{B}}^i \boldsymbol{h}_{\mathcal{B}}^{i\,H}\right)^{-1} \boldsymbol{h}_{\mathcal{S}}^i \right], \quad \mathcal{S} \subseteq \mathcal{A}. \quad (32)$$

The following lemma is readily verified using the properties of a polymatroid.

**Lemma 5** $\theta(\boldsymbol{h}^i, \mathcal{A}, \mathcal{B}, \boldsymbol{R}^{\min}, \boldsymbol{\rho})$ *can be computed using*

$$\theta(\boldsymbol{h}^i, \mathcal{A}, \mathcal{B}, \boldsymbol{R}^{\min}, \boldsymbol{\rho}) = \min_{S \neq \emptyset, S \subseteq \mathcal{A}} \frac{\Delta(\boldsymbol{h}^i, S, \mathcal{B}, \boldsymbol{R}^{\min})}{\sum_{j \in S} \rho_j},$$

*where*

$$\Delta(\boldsymbol{h}^i, S, \mathcal{B}, \boldsymbol{R}^{\min}) = \log \det \left[ \boldsymbol{I} + \boldsymbol{h}_{S}^{i\,H} \left(1 + \boldsymbol{h}_{\mathcal{B}}^i \boldsymbol{h}_{\mathcal{B}}^{i\,H}\right)^{-1} \boldsymbol{h}_{S}^i \right] - \sum_{j \in S} R_j^{\min}.$$

Therefore, the maximum rate increment factor for sustaining weighted max-min fairness such that the $i^{th}$ user is decodable is given by

$$\theta_i^* \triangleq \max_{\mathcal{G} \subseteq \mathcal{K}, i \in \mathcal{G}} \theta(\boldsymbol{h}^i, \mathcal{G}, \mathcal{K}\backslash\mathcal{G}, \boldsymbol{R}^{\min}, \boldsymbol{\rho}). \tag{33}$$

Based on (33) a direct (naive) way to compute $\theta_i^*$ is to exhaustively search all the partitions $\{\mathcal{G}, \mathcal{K}\backslash\mathcal{G}\}$ such that $i \in \mathcal{G}$. Such an exhaustive search has a complexity that scales exponentially as $O(3^{M_s})$. Our contribution is to propose an efficient algorithm that finds $\theta_i^*$ for each user $i$ (along with an optimal partition), with a complexity that is polynomial in $M_s$. Before we proceed further, we note that $\Delta(\cdot)$ has the following properties.

$$\mathcal{A}, \mathcal{B}, \mathcal{C} \text{ disjoint} \Rightarrow \Delta(\boldsymbol{h}^i, \mathcal{A}, \mathcal{B} \cup \mathcal{C}, \boldsymbol{R}^{\min}) + \Delta(\boldsymbol{h}^i, \mathcal{B}, \mathcal{C}, \boldsymbol{R}^{\min}) = \Delta(\boldsymbol{h}^i, \mathcal{A} \cup \mathcal{B}, \mathcal{C}, \boldsymbol{R}^{\min}), \tag{34}$$

$$\text{and for any } \mathcal{B} \subseteq \mathcal{C} \Rightarrow \Delta(\boldsymbol{h}^i, \mathcal{A}, \mathcal{B}, \boldsymbol{R}^{\min}) \geq \Delta(\boldsymbol{h}^i, \mathcal{A}, \mathcal{C}, \boldsymbol{R}^{\min}). \tag{35}$$

Now for the $i^{th}$ secondary receiver we introduce rate increments $\{r_1^i, \ldots, r_{M_s}^i\}$ where $r_k^i$ is the rate increment for the $k^{th}$ user such that the $i^{th}$ user remains decodable at receiver $i$ and weighted max-min fairness is sustained, i.e., $\min_k\{\frac{r_k^i}{\rho_k}\} = \theta_i^*$. Algorithm 3, provided in the sequel, is a computationally efficient scheme for finding the set of rate increments $\{r_1^i, \ldots, r_{M_s}^i\}$ for each given user $i$. We note that since the function $\Delta(\cdot)$ is sub-modular, using sub-modular function minimization techniques [16] we can compute $\delta^k$ in step 3 of Algorithm 3 with a complexity that is polynomial in $|\mathcal{S}|$. Consequently, since the number of loops in Algorithm 3 can be no greater than $M_s$, Algorithm 3 itself has a complexity that is polynomial in $M_s$. We now prove the optimality of Algorithm 3.

**Theorem 2** *For a given $\boldsymbol{R}^{\min}$ such that the $i^{th}$ user is decodable at receiver $i$, the $i^{th}$ user is also decodable under the rate vector $\{R_k^{\min} + r_k^i\}_{k=1}^{M_s}$, where $\{r_k^i\}$ is yielded by Algorithm 3. Further,*

$$\min_{k \in \mathcal{K}} \frac{r_k^i}{\rho_k} \geq \min_{k \in \mathcal{K}} \frac{\tilde{r}_k^i}{\rho_k}$$

*where $\{\tilde{r}_1^i, \ldots, \tilde{r}_{M_s}^i\}$ is any other rate increment vector for which the $i^{th}$ user is decodable at receiver $i$ under the rates $\{R_k^{\min} + \tilde{r}_k^i\}_{k=1}^{M_s}$.*

*Proof:* See [17, Theorem 2]. ∎

In Algorithm 3, user $i$ makes rate increment suggestions for all users (including itself) denoted by $\{r_1^i, \ldots, r_{M_s}^i\}$ such that the updated rate vector remains decodable at receiver $i$. Based on such rate increment suggestions by all users, we construct Algorithm 4 to determine the rate increment for each secondary user. Algorithm 4 is an iterative algorithm and in each iteration each user $j$ receives $M_s$ rate increment suggestions from all users which are denoted by $\{r_j^1, \ldots, r_j^{M_s}\}$. After receiving all such rate increment suggestions, the $j^{th}$ user picks the smallest rate increment suggested for it, i.e., $\min_{1 \leq i \leq M_s} r_j^i$. In the following theorem we prove the optimality properties of Algorithm 4.

**Theorem 3** *The distributed (iterative) weighted max-min fair rate allocation algorithm has the following properties:*

1. *It is monotonic in the sense that $\boldsymbol{R}^{(q+1)} \succeq \boldsymbol{R}^{(q)}$ and is convergent.*

2. *At each iteration the vector $\boldsymbol{R}^{(q)}$ is max-min optimal, i.e., for any other arbitrary decodable rate vector $\tilde{\boldsymbol{R}} \succeq \boldsymbol{R}^{\min}$ we have*

$$\min_{k \in \mathcal{K}} \frac{R_k^{(q)} - R_k^{\min}}{\rho_k} \geq \min_{k \in \mathcal{K}} \frac{\tilde{R}_k - R_k^{\min}}{\rho_k}, \quad \forall \, q \geq 1.$$

3. *The rate allocation $\boldsymbol{R}^*$ yielded by Algorithm 4 is also pareto-optimal, i.e., for any arbitrary decodable rate vector $\tilde{\boldsymbol{R}} \succeq \boldsymbol{R}^{\min}$ such that $\tilde{R}_k > R_k^*$ for some $k \in \mathcal{K}$, we must have that $\exists \, j \neq k : \tilde{R}_j < R_j^*$.*

*Proof:* See Appendix E ∎

Note that after one iteration of Algorithm 4, we can obtain another optimal rate allocation that satisfies a stricter notion of fairness. In particular, letting $\hat{x} = \min_{k \in \mathcal{K}} \frac{R_k^{(1)} - R_k^{\min}}{\rho_k}$ with $\hat{\boldsymbol{R}} = \boldsymbol{R}^{\min} + \hat{x} \cdot \boldsymbol{\rho}$ we have that $\hat{\boldsymbol{R}}$ is decodable and satisfies

$$\hat{\boldsymbol{R}} \succeq \tilde{\boldsymbol{R}},$$

where $\tilde{\boldsymbol{R}}$ is any decodable rate-vector such that $\tilde{\boldsymbol{R}} = \boldsymbol{R}^{\min} + x \cdot \boldsymbol{\rho}$ for some $x \geq 0$.

## 5.2 Maximum Likelihood Decoders

In order to address the case when the MLD is employed at each secondary receiver, we provide Algorithm 4MLD which can be initialized with any rate vector $\boldsymbol{R}^{\min}$ that is decodable when the MLD is employed at each receiver. The optimality of Algorithm 4MLD is stated in the following theorem. The proof of the theorem follows along similar lines as the one given for Theorem 3 and hence is skipped for brevity.

**Theorem 4** *The distributed (iterative) weighted max-min fair rate allocation algorithm for the MLD has the following properties:*

1. It is monotonic in the sense that $\boldsymbol{R}^{(q+1)} \succeq \boldsymbol{R}^{(q)}$ and is convergent.

2. At each iteration the vector $\boldsymbol{R}^{(q)}$ is max-min optimal, i.e., for any other arbitrary rate vector $\tilde{\boldsymbol{R}} \succeq \boldsymbol{R}^{\min}$ that is decodable using the MLD at each receiver, we have

$$\min_{k \in \mathcal{K}} \frac{R_k^{(q)} - R_k^{\min}}{\rho_k} \geq \min_{k \in \mathcal{K}} \frac{\tilde{R}_k - R_k^{\min}}{\rho_k}, \quad \forall\, q \geq 1.$$

3. The rate allocation $\boldsymbol{R}^{\mathrm{ML}}$ yielded by Algorithm 4MLD is also pareto-optimal, i.e., for any arbitrary rate vector $\tilde{\boldsymbol{R}} \succeq \boldsymbol{R}^{\min}$ decodable using the MLD at each receiver, such that $\tilde{R}_k > R_k^{\mathrm{ML}}$ for some $k \in \mathcal{K}$, we must have that $\exists\, j \neq k : \tilde{R}_j < R_j^{\mathrm{ML}}$.

## 6 Simulation Results

In this section, we provide simulation results to assess the performance of different beamforming designs and rate allocation schemes. For convenience, throughout all the simulations we consider the setting $\alpha_i = \rho_i = 1$ (so that $\gamma_i = 2$) for $i = 1, \ldots, M_s$, with $\beta_j = 5$ for $j = 1, \ldots, M_p$.

In figures 1 and 2 we examine the rate optimization problem $\mathcal{R}(P_0)$ for the MMSE and the UGD receivers. For this purpose we consider a cognitive radio network with three secondary transceivers ($M_s = 3$) and two primary transceivers ($M_p = 2$). Each primary and secondary user has three transmit ($N_s = N_p = 3$) and one receive antenna, respectively. The sum power available for all secondary transmitters is taken to be $P_0 = 20$ dB. We first obtain the optimal beamforming vectors using Algorithm 2. We then implement Algorithm 4 (with four iterations and identical priorities for all users) after initializing it with $\boldsymbol{R}^{\min} = \boldsymbol{R}^{\mathrm{MMSE}}$ to obtain the optimal rate increments. In Fig. 1, we consider 20 channel realizations (generated assuming all fading coefficients to be i.i.d. complex Gaussian) and for each realization, we plot the normalized minimum secondary rate obtained using the optimal beam vectors and the MMSE decoder at each secondary receiver, where the minimum rate is normalized by the minimum rate obtained after Algorithm 4. We also compare these results with the performance of MMSE receivers when the beamformers are obtained through channel matching, that is, the beamforming vector used by each secondary transmitter is aligned with channel direction to its intended receiver. The results for the latter case are also normalized by the minimum rates obtained after deploying Algorithm 4 with the optimized beamformers. Fig. 2 considers the same setup, but instead shows the normalized *sum-rates*. These two plots demonstrate that designing the beamformers through Algorithm 2 brings about considerable improvement in the minimum-rate and sum-rate when compared against the channel matching design. Further rate gains are achievable at the expense of additional signaling overhead and advanced decoding at each receiver.

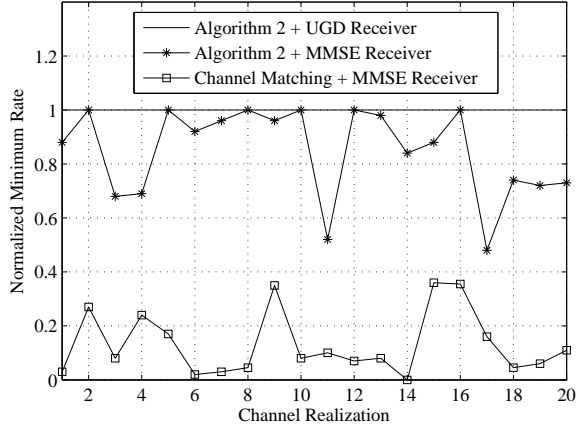

Figure 1: UGD vs. MMSE: the normalized minimum rates achieved with the sum-power constraint 20 dB.

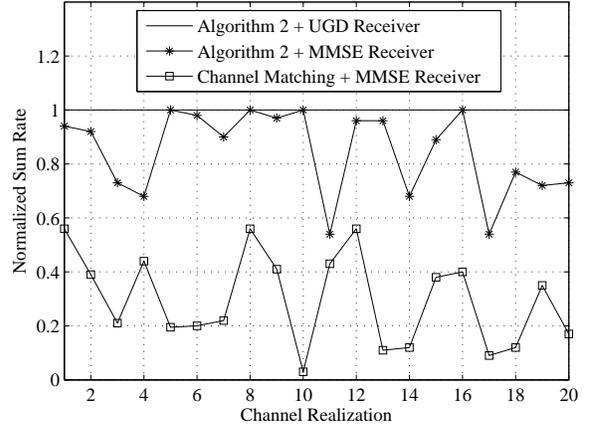

Figure 2: UGD vs. MMSE: the normalized sum-rates achieved with the sum-power constraint 20 dB.

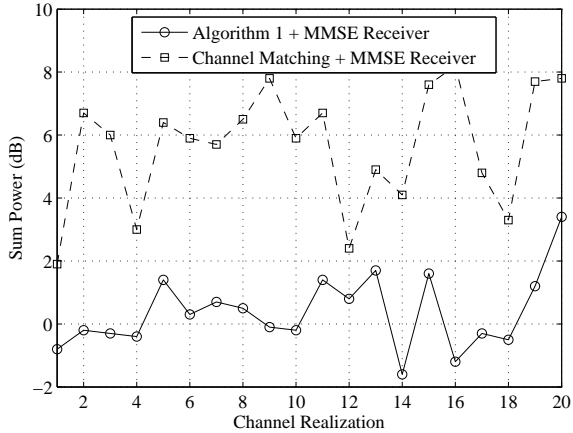

Figure 3: MMSE receiver: sum-power required by Algorithm 1 vs. channel matching for achieving $\mathsf{SINR}_i \geq 2, \forall\, i$

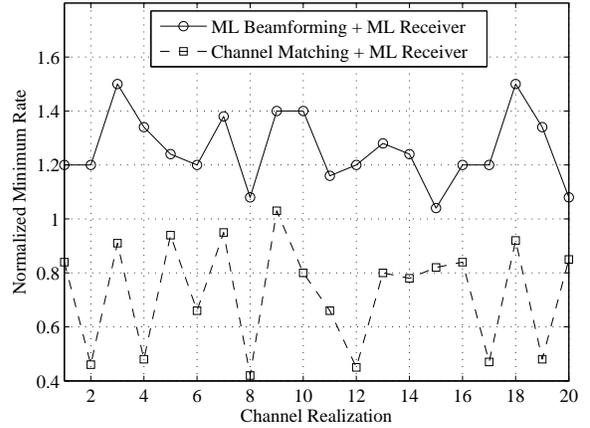

Figure 4: ML receiver: normalized minimum rates achieved using optimized beamformers (Section IV-B) vs. channel matching with the sum-power constraint of 10 dB.

In Fig. 3 we consider the power optimization problem $\mathcal{P}(\boldsymbol{\gamma})$ for the MMSE receivers. With the constraints of achieving $\mathsf{SINR}_i \geq 2 \,\forall\, i$, we compare the sum-power required when the beamformers are obtained through Algorithm 1 and when they are obtained through channel matching. In particular, we assume a cognitive radio network with four primary transceivers ($M_p = 4$) and three secondary transceivers ($M_s = 3$), each equipped with four transmit and one receive antenna. For each realization we plot the secondary sum power when the secondary transmitters employ optimal beams obtained using Algorithm 1. Also plotted is the secondary sum power when the secondary transmitters employ beams that are matched to the forward channel vectors to their respective intended receivers and which are then scaled subject to the SINR and margin constraints. Note that a careful design of the beamforming vectors can result in substantial power savings.

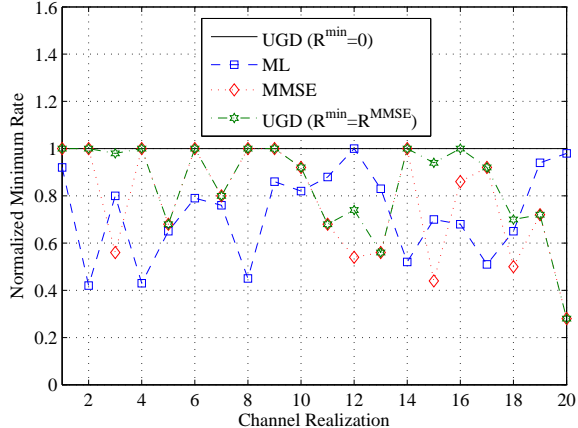 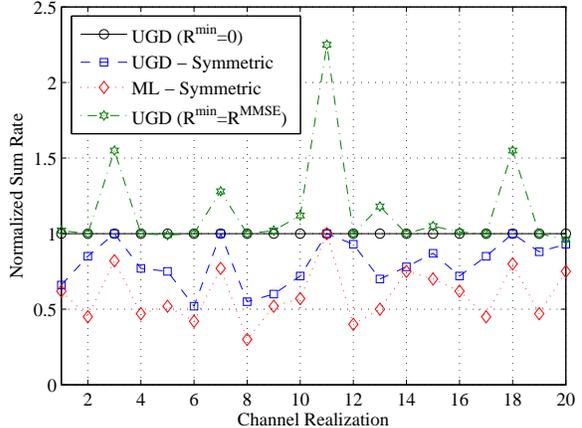

Figure 5: Comparing the normalized minimum rate yielded by different receivers for a given beamformer design.

Figure 6: Comparing the normalized sum-rate yielded by different receivers for a given beamformer design.

In Fig. 4 we consider the rate optimization problem $\mathcal{R}(P_0)$ when each secondary receiver employs ML decoding. We assume the same network setup as that in Fig. 3 and set the secondary sum power budget to be 10 dB. For each realization we plot the minimum secondary rates when the secondary transmitters employ optimized beams obtained using the method described in Section 4.2. As a comparison baseline, we have also plotted the minimum secondary rates when the secondary transmitters employ beamformers obtained via channel matching. The gains yielded by the optimized beamforming design is evident from the figure.

Finally, in figures 5 and 6 we demonstrate the relative gains of the different rate allocation algorithms and the tradeoff between system efficiency and fairness. We assume that a set of beam vectors based on the channel matching scheme is given for each channel realization. In Fig. 5 we plot the normalized minimum secondary rates obtained when each secondary receiver employs the MMSE decoder, the ML decoder and the UGD, respectively. The minimum rates for the ML decoder and the UGD were determined using Algorithms 4MLD and 4, respectively. The Algorithm 4MLD is initialized by setting $\boldsymbol{R}^{\min} = \boldsymbol{0}$. For initializing Algorithm 4 we have considered two cases with $\boldsymbol{R}^{\min} = \boldsymbol{0}$ and $\boldsymbol{R}^{\min} = \boldsymbol{R}^{\text{MMSE}}$, where $\boldsymbol{R}^{\text{MMSE}}$ is the vector of rates yielded by the MMSE receivers. All rates are normalized by that obtained for the UGD with $\boldsymbol{R}^{\min} = \boldsymbol{0}$. Note that the minimum rate yielded by the UGD with $\boldsymbol{R}^{\min} = \boldsymbol{0}$ is provably larger than those achieved by the ML and MMSE decoders as well as the one yielded by the UGD but with $\boldsymbol{R}^{\min} = \boldsymbol{R}^{\text{MMSE}}$. Thus the minimum rate yielded by the UGD with $\boldsymbol{R}^{\min} = \boldsymbol{0}$ is max-min optimal. Further, the minimum rate achieved by the ML decoder is seen to be sometimes smaller (albeit not always) than the one obtained by MMSE decoder.

In Fig. 6 we plot the corresponding normalized sum rates. In particular, we plot the normalized sum rates obtained for the ML decoder and the UGD (denoted by UGD-sym) under the stricter notion of fairness that all users are assigned identical rate. We remark that even under this notion of fairness the rate vector obtained for the UGD *has better efficiency than the one obtained for the ML decoder*. The rates are normalized using the sum rate obtained for the UGD using Algorithm 4 (denoted by UGD), with four iterations and after initializing it with $\boldsymbol{R}^{\min} = \boldsymbol{0}$. Note that the latter rate vector has an identical value of its minimum element but a larger sum compared to the one corresponding to UGD-sym since we do not insist that all elements have an identical value. Finally, also plotted is the normalized sum rate obtained for the UGD using Algorithm 4 (denoted by UGD-MMSE), again with four iterations but after initializing it with $\boldsymbol{R}^{\min} = \boldsymbol{R}^{\text{MMSE}}$. Note that the rate vectors corresponding to UGD and UGD-sym, respectively, have a larger minimum rate than the one corresponding to UGD-MMSE but the latter one significantly improves the sum-rate (efficiency).

# 7 Conclusions

Our focus was on decentralized multi-antenna cognitive radio networks where secondary transceivers co-exist with primary ones. We devised distributed algorithms for optimal beamforming and rate allocation in such networks. We formulated the optimization problems for the cases when the secondary receivers employ single-user decoders, maximum likelihood decoders and unconstrained group decoders. An optimal distributed algorithm is obtained for the case when each secondary receiver employs single-user decoding. The algorithm is optimal in the sense that it maximizes the minimum weighted rate subject to a weighted sum power budget for the secondary users and interference margin constraints imposed by the primary users. We also obtained a centralized sub-optimal algorithm for the case when each secondary receiver employs the maximum likelihood decoder. Finally, for advanced decoders at the secondary receivers, we proposed distributed low-complexity fair rate allocation algorithms to boost the system efficiency and maintain a notion of fairness.

## A  Proof of Theorem 1

Note that since the optimization problem corresponding to $\mathcal{R}(P_0)$ is always feasible, we conclude that there exists a set of beamformers such that the rate vector $\mathcal{R}(P_0) \cdot \boldsymbol{\rho}$ is decodable, the margin constraints are satisfied and the weighted sum power does not exceed $P_0$. Consequently, we conclude that $\mathcal{P}(\mathcal{R}(P_0) \cdot \boldsymbol{\rho}) \leq P_0$. Next, suppose the optimization problem corresponding to $\mathcal{P}(\boldsymbol{\rho})$ is feasible and let $\{\boldsymbol{w}_i^*\}$ be the optimal set of beamformers with $\sum_i \alpha_i \|\boldsymbol{w}_i^*\|^2 = \mathcal{P}(\boldsymbol{\rho})$. Then since for the beamformer design $\{\boldsymbol{w}_i^*\}$ the rate vector $\boldsymbol{\rho}$ is decodable and the margin constraints are satisfied, we must have that $\mathcal{R}(\mathcal{P}(\boldsymbol{\rho})) \geq 1$. On the other hand, $\mathcal{R}(\mathcal{P}(\boldsymbol{\rho}))$ cannot exceed 1. If it does, using the fact that $\mathcal{R}(P_0)$ is continuous and non-decreasing in $P_0$, we can conclude that there exists a set of beamformers $\{\tilde{\boldsymbol{w}}_i\}$ with $\sum_{i=1}^{M_s} \alpha_i \|\tilde{\boldsymbol{w}}_i\|^2 = P_0 < \mathcal{P}(\boldsymbol{\rho})$ and at the same time $\mathcal{R}(P_0) = 1$. This means that $\boldsymbol{\rho}$ is decodable and the margin constraints are satisfied which contradicts the optimality of $\mathcal{P}(\boldsymbol{\rho})$ and hence we must have $\mathcal{R}(\mathcal{P}(\boldsymbol{\rho})) = 1$.

## B  Proof of Lemma 2

We first find an upper bound on the term in (11) as follows.

$$\min_i \frac{|[\boldsymbol{QT}]_{i,i}|^2}{\sum_{j \neq i} |[\boldsymbol{QT}]_{i,j}|^2 + \boldsymbol{\sigma}(i)} \leq \min_i \frac{|[\boldsymbol{QT}]_{i,i}|^2}{\sum_j |[\boldsymbol{QT}]_{i,j}|^2 - |[\boldsymbol{QT}]_{i,i}|^2} = \min_i \left(\frac{[\boldsymbol{QTT}^H\boldsymbol{Q}^H]_{i,i}}{|[\boldsymbol{QT}]_{i,i}|^2} - 1\right)^{-1}$$
$$= \left(\max_i \frac{[\boldsymbol{QTT}^H\boldsymbol{Q}^H]_{i,i}}{|[\boldsymbol{QT}]_{i,i}|^2} - 1\right)^{-1} \leq \left(\frac{1}{M_s + M_p} \sum_i \frac{[\boldsymbol{QTT}^H\boldsymbol{Q}^H]_{i,i}}{|[\boldsymbol{QT}]_{i,i}|^2} - 1\right)^{-1}. \quad (36)$$

Assuming the singular value decomposition (SVD) $\boldsymbol{QT} = \boldsymbol{U\Lambda V}^H$ such that $\boldsymbol{u}_i$ and $\boldsymbol{v}_i$ are the $i^{th}$ rows of the unitary matrices $\boldsymbol{U}$ and $\boldsymbol{V}$, respectively, we get

$$\frac{[\boldsymbol{QTT}^H\boldsymbol{Q}^H]_{i,i}}{|[\boldsymbol{QT}]_{i,i}|^2} = \frac{\boldsymbol{u}_i\boldsymbol{\Lambda\Lambda}^H\boldsymbol{u}_i^H}{|\boldsymbol{u}_i\boldsymbol{\Lambda v}_i^H|^2} \geq \frac{1}{\boldsymbol{v}_i\boldsymbol{v}_i^H} = \frac{1}{[(\boldsymbol{QT})^\dagger\boldsymbol{QT}]_{i,i}}, \quad (37)$$

where $()^\dagger$ denotes the pseudo-inverse operator and the inequality above holds due to Cauchy-Schwartz inequality. Considering (36) and (37) and using the harmonic mean-arithmetic mean inequality we get

$$\min_i \frac{|[\boldsymbol{QT}]_{i,i}|^2}{\sum_{j\neq i}|[\boldsymbol{QT}]_{i,j}|^2 + \boldsymbol{\sigma}(i)} \leq \left(\frac{1}{M_s + M_p}\sum_i \frac{1}{[(\boldsymbol{QT})^\dagger \boldsymbol{QT}]_{i,i}} - 1\right)^{-1}$$
$$\leq \left(\frac{M_s + M_p}{\sum_i [(\boldsymbol{QT})^\dagger \boldsymbol{QT}]_{i,i}} - 1\right)^{-1} = \left(\frac{M_s + M_p}{\text{tr}[(\boldsymbol{QT})^\dagger \boldsymbol{QT}]} - 1\right)^{-1}$$
$$= \left(\frac{M_s + M_p}{\text{rank}[\boldsymbol{QT}]} - 1\right)^{-1} \leq \left(\frac{M_s + M_p}{\text{rank}[\boldsymbol{Q}]} - 1\right)^{-1}. \tag{38}$$

Therefore, to satisfy all the SINR constraints it is necessary that $\left(\frac{M_s+M_p}{\text{rank}[\boldsymbol{Q}]} - 1\right)^{-1} \geq 1$.

## C Proof of Lemma 3

By rearranging the constraints in (11) and using the definitions of $\boldsymbol{Q}$ and $\boldsymbol{T}$ we have

$$2|[\boldsymbol{QT}]_{i,i}|^2 \geq \sum_{j=1}^{M_s+M_p} |[\boldsymbol{QT}]_{i,j}|^2 + \boldsymbol{\sigma}(i) = \left\|\begin{array}{c}(\boldsymbol{QT})^T \boldsymbol{e}_i \\ \sqrt{\boldsymbol{\sigma}(i)}\end{array}\right\|^2 \quad \text{for } i = 1, \ldots, M_s + M_p,$$

where $\boldsymbol{e}_i$ is a column vector of length $M_s + M_p$ with its $i^{th}$ element 1 and the rest 0. As the angular rotation of the beamforming vectors with any arbitrary phase does not affect either the objective or any of the constraints of $\tilde{\mathcal{P}}(\boldsymbol{\gamma})$, therefore we confine the solution to satisfy $[\boldsymbol{QT}]_{i,i} \geq 0$ for $i = 1, \ldots, M_s$. Also note that for $j > M_s$, $[\boldsymbol{QT}]_{i,j}$ is a fixed value. Hence, $\tilde{\mathcal{P}}(\boldsymbol{\gamma})$ is equivalent to

$$\begin{cases} \min_{\{\tilde{\boldsymbol{w}}_i^s\}} & \sum_{i=1}^{M_s} \|\tilde{\boldsymbol{w}}_i^s\|^2 \\ \text{s.t.} & \begin{bmatrix} \sqrt{2}|[\boldsymbol{QT}]_{i,i}| \\ (\boldsymbol{QT})^T \boldsymbol{e}_i \\ \sqrt{\boldsymbol{\sigma}(i)} \end{bmatrix} \succeq_\kappa 0, \quad \text{for } i = 1, \ldots, M_s + M_p \end{cases} \tag{39}$$

Since for $i > M_s$ we have $|[\boldsymbol{QT}]_{i,i}| = [\boldsymbol{QT}]_{i,i} = 1$, the constraints in (39) are all second order cones. Therefore since the objective function is convex, (39) is a convex problem for which strong duality holds and the primal and the Lagrangian dual problems achieve the same optimal value. The Lagrangian dual of (39) is given by

$$L_1\left(\{\tilde{\boldsymbol{w}}_i^s\}, \boldsymbol{D}\right) = \sum_{i=1}^{M_s} \|\tilde{\boldsymbol{w}}_i^s\|^2 + \sum_{i=1}^{M_s+M_p} d_i \left(\left\|\begin{array}{c}(\boldsymbol{QT})^T \boldsymbol{e}_i \\ \sqrt{\boldsymbol{\sigma}(i)}\end{array}\right\| - \sqrt{2}[\boldsymbol{QT}]_{i,i}\right), \tag{40}$$

where $\boldsymbol{D} = \text{diag}(d_1, \ldots, d_{M_s+M_p})$ and the Lagrangian dual objective function is

$$g_1(\boldsymbol{D}) = \min_{\{\tilde{\boldsymbol{w}}_i^s\}} L_1\left(\{\tilde{\boldsymbol{w}}_i^s\}, \boldsymbol{D}\right). \tag{41}$$

Because of the convexity of (39) we have $\tilde{\mathcal{P}}(\boldsymbol{\gamma}) = \max_{\boldsymbol{D} \succeq 0} g_1(\boldsymbol{D})$. Defining

$$q_i \triangleq \left\| \frac{(\boldsymbol{QT})^T \boldsymbol{e}_i}{\sqrt{\boldsymbol{\sigma}(i)}} \right\| + \sqrt{2}[\boldsymbol{QT}]_{i,i}, \tag{42}$$

along with some manipulations on $L_1(\cdot, \cdot)$ yields

$$\begin{aligned}
L_1\left(\{\tilde{\boldsymbol{w}}_i^s\}, \boldsymbol{D}\right) &= \sum_{i=1}^{M_s} \|\tilde{\boldsymbol{w}}_i^s\|^2 + \sum_{i=1}^{M_s+M_p} \frac{d_i}{q_i} \left( \left\| \frac{(\boldsymbol{QT})^H \boldsymbol{e}_i}{\sqrt{\boldsymbol{\sigma}(i)}} \right\|^2 - 2[\boldsymbol{QT}]_{i,i}^2 \right) \\
&= \sum_{i=1}^{M_s} \|\tilde{\boldsymbol{w}}_i^s\|^2 + \sum_{i=1}^{M_s} \frac{d_i}{q_i} \left( \sum_{j \neq i} |\tilde{\boldsymbol{h}}_{i,j}^{s,s} \tilde{\boldsymbol{w}}_j^s|^2 + \sum_j |\boldsymbol{h}_{i,j}^{s,p} \boldsymbol{w}_j^p|^2 + \sigma_i^s - |\tilde{\boldsymbol{h}}_{i,i}^{s,s} \tilde{\boldsymbol{w}}_i^s|^2 \right) \\
&\quad + \sum_{i=1}^{M_p} \frac{d_{M_s+i}}{q_{M_s+i}} \left( \sum_j |\tilde{\boldsymbol{h}}_{i,j}^{p,s} \tilde{\boldsymbol{w}}_j^s|^2 - 1 \right).
\end{aligned} \tag{43}$$

Note that the $\{q_i\}$ as defined in (42) are strictly positive. As a result, we may replace the Lagrangian coefficients with $\varphi_i \triangleq \frac{d_i}{q_i}$ for $i = 1, \ldots, M_s + M_p$ and upon re-writing (43) we can readily verify it to be identical to the Lagrangian dual of (10) having the dual variables $\{\varphi_j\}$. Therefore, by taking into account that (10) and (39) are equivalent and their Lagrangian duals are the same, they should exhibit identical duality gap. On the other hand, (39) is convex and satisfies Slater's condition and thus has a zero duality gap. As a result, (10) exhibits a zero duality gap too.

## D  Proof of Lemma 4

Consider any rate vector $\boldsymbol{R}$ satisfying the condition in (19). Let $\epsilon > 0$ be arbitrarily fixed and let $\boldsymbol{C}$ be the ensemble of multi-user codebooks corresponding to any rate vector $\boldsymbol{R}' \prec \boldsymbol{R}$. Define any multi-user codebook (i.e., a set of $M_s$ codes, one for each secondary user) by $C \in \boldsymbol{C}$ and define $E_i(C)$ as the error event for user $i$ (at its designated receiver) when the multi-user codebook $C$ is employed. We assume that the multi-user codebooks are picked using the product Gaussian measure, i.e., each coordinate of each codeword of each user is generated independently using a Gaussian distribution with zero-mean and unit variance. Our objective is to show that there exists at-least one multi-user codebook $C$ for which $\Pr(E_i(C)) \leq \epsilon$, $\forall\, i$.

For this purpose, we consider the term $\Pr(\cap_{i=1}^{K}\{\Pr(E_i(C)) \leq \epsilon\})$, where the outer probability is over the set $C$ and the inner probability is over the set of all noise realizations. We obtain the following bound

$$\Pr\left(\cap_{i=1}^{M_s}\{\Pr(E_i(C)) \leq \epsilon\}\right) = 1 - \Pr\left(\cup_{i=1}^{M_s}\{\Pr(E_i(C)) > \epsilon\}\right) \geq 1 - \sum_{i=1}^{M_s}\Pr(\Pr(E_i(C)) > \epsilon). \quad (44)$$

Using Markov's inequality, we obtain

$$1 - \sum_{i=1}^{M_s}\Pr(\Pr(E_i(C)) > \epsilon) \geq 1 - \sum_{i=1}^{M_s}\frac{\mathbb{E}[\Pr(E_i(C))]}{\epsilon}.$$

Next at any receiver $i$, the event $E_i(C)$ is the union of $2^{M_s-1}$ disjoint events, where the $k^{th}$ event is the event that errors occur only for all users in the set $\mathcal{V}_i^k$, for $k = 1, \cdots, 2^{M_s-1}$, respectively. Then since the rate vector $\boldsymbol{R}$ satisfies the condition in (19), it can be verified using the random coding upper bounds and the techniques developed in [13] that the terms $\{\mathbb{E}[\Pr(E_i(C))]\}_{i=1}^{M_s}$ can be made arbitrarily small. Thus, for sufficiently long codeword lengths the term $\Pr(\cap_{i=1}^{K}\{\Pr(E_i(C)) \leq \epsilon\})$ is bounded away from zero and hence we can conclude that exists at-least one multi-user codebook $C$ satisfying the power constraints, for which $\Pr(E_i(C)) \leq \epsilon$, $\forall\, i$.

## E  Proof of Theorem 3

**Claim 1:**

Since $\boldsymbol{R}^{\min}$ is decodable, as an straightforward application of Theorem 2 we find that $\boldsymbol{R}^{(1)}$ is also decodable and $\boldsymbol{R}^{(1)} \succeq \boldsymbol{R}^{\min}$. In general, at the $(q+1)^{th}$ iteration for finding the rate vector $\boldsymbol{R}^{(q+1)}$ we have set $\boldsymbol{R}^{\min} = \boldsymbol{R}^{(q)}$ and again by using Theorem 2 we conclude that $\boldsymbol{R}^{(q+1)}$ is decodable and $\boldsymbol{R}^{(q+1)} \succeq \boldsymbol{R}^{(q)}$. Finally, as the set of rate vectors $\{\boldsymbol{R}^{(q+1)}\}$ is monotonically increasing and the set of decodable rate vectors is bounded, the convergence is guaranteed.

**Claim 2:**

By invoking $\boldsymbol{R}^{(q)} \succeq \cdots \succeq \boldsymbol{R}^{(1)}$ from the first part we get

$$\min_{k \in \mathcal{K}} \frac{R_k^{(q)} - R_k^{\min}}{\rho_k} \geq \cdots \geq \min_{k \in \mathcal{K}} \frac{R_k^{(1)} - R_k^{\min}}{\rho_k}. \quad (45)$$

Now, for the given rate vector $\tilde{\boldsymbol{R}}$ let us define $\tilde{r}_k^i \triangleq \tilde{R}_k - R_k^{\min}$ for $i = 1, \ldots, M_s$. By noting that $R^{(1)} = R_k^{\min} + \min_{1 \leq i \leq M_s}\{r_k^i\}$, where $\{r_k^i\}$ are the rate recommendations made after the first iteration, we get

$$\min_{k \in \mathcal{K}} \frac{R_k^{(1)} - R_k^{\min}}{\rho_k} = \min_{k \in \mathcal{K}} \frac{\min_{1 \leq i \leq M_s}\{r_k^i\}}{\rho_k} = \min_{1 \leq i \leq M_s} \min_{k \in \mathcal{K}} \frac{r_k^i}{\rho_k} \geq \min_{1 \leq i \leq M_s} \min_{k \in \mathcal{K}} \frac{\tilde{r}_k^i}{\rho_k} \quad (46)$$

$$= \min_{k \in \mathcal{K}} \min_{1 \leq i \leq M_s} \frac{\tilde{r}_k^i}{\rho_k} = \min_{k \in \mathcal{K}} \min_{1 \leq i \leq M_s} \frac{R_k - R_k^{\min}}{\rho_k} = \min_{k \in \mathcal{K}} \frac{R_k - R_k^{\min}}{\rho_k}, \quad (47)$$

where (46) holds due to Theorem 2. By putting together (45) and (47) the desired result is established.

**Claim 3:**

We consider the output of Algorithm 4 and show that for this rate allocation, any increase in the rate of any user will incur a decrease in the rate of some other user and thus, $\boldsymbol{R}^*$ is the pareto-optimal solution. For this purpose, we investigate the possibility of increasing the rate of a specific user while keeping those of the others' unchanged. Without loss of generality we examine whether the rate vector $\tilde{\boldsymbol{R}} = \{R_1^* + \varepsilon, R_2^*, \ldots, R_{M_s}^*\}$ is decodable for some $\varepsilon > 0$.

At each iteration, each specific user receives rate increment suggestions by all other users among which the user with the lowest rate increment suggestion identifies the rate increment for that specific user. At the final iteration, let us assume that the lowest rate increment recommendation for user 1 is made by the $i^{th}$ receiver, i.e., $r_1^i = \min_j\{r_1^j\} = 0$. Also, let $\{\mathcal{B}^1, \ldots, \mathcal{B}^m, \mathcal{B}^{m+1}, \ldots, \mathcal{B}^p\}$ denote the sets found at the $i^{th}$ receiver by Algorithm 3 in the last iteration of Algorithm 4, using $\boldsymbol{R}^*$ as the minimum rate vector and denote their respective corresponding values by $\{\delta^1, \ldots, \delta^m, \delta^{m+1}, \ldots, \delta^p\}$. Suppose $i \in \mathcal{B}^{m+1}$ and since the $i^{th}$ user must be decodable, we must have $\delta^{m+1} \geq 0$. Using the arguments employed in the proof of [18, Theorem2], we can show that $\delta^1 \leq \delta^2 \leq \cdots \leq \delta^p$. Based on this observation we can deduce the following properties for the sets $\{\mathcal{B}^i\}$ and $\{\delta^i\}$:

1. $\delta^{m+2} > 0$: Clearly when $\delta^{m+1} > 0$ we must have $\delta^{m+2} > 0$. Now suppose $\delta^{m+1} = 0$ so that $\delta^{m+2} \geq 0$. Assume $\delta^{m+2} = 0$. Then, since $\Delta(\boldsymbol{h}^i, \mathcal{B}^{m+2}, \cup_{j=1}^{m+1}\mathcal{B}^j, \boldsymbol{R}^*) = \Delta(\boldsymbol{h}^i, \mathcal{B}^{m+1}, \cup_{j=1}^{m}\mathcal{B}^j, \boldsymbol{R}^*) = 0$ it can be shown that $\Delta(\boldsymbol{h}^i, \mathcal{B}^{m+2}, \cup_{j=1}^{m}\mathcal{B}^j, \boldsymbol{R}^*) = 0$. This is a contradiction since it implies that in Algorithm 3, line 3, we could have chosen $\mathcal{B}^{m+2}$ instead of $\mathcal{B}^{m+1}$. Thus, $\delta^{m+2} > \delta^{m+1} \geq 0$.

2. $1 \in \mathcal{B}^{m+1}$: First, $1 \notin \mathcal{B}^j$ for $j \leq m$ since otherwise the $i^{th}$ user would recommend $r_1^i = +\infty$ which is a contradiction. On the other hand, if $i \in \mathcal{B}^j$ for $j \geq m+2$ then the $i^{th}$ user would recommend the rate increment $\delta^j \rho_1 > 0$ which is also a contradiction.

3. $\delta^{m+1} = 0$: Since $1 \in \mathcal{B}^{m+1}$, due to $\boldsymbol{R}^*$ being the convergence point, $\delta^{m+1}$ cannot be greater than zero as otherwise it leads to a higher rate for the $1^{st}$ user. By taking into account that $\delta^{m+1} \geq 0$ we get $\delta^{m+1} = 0$.

Recall that we have $\delta^1 \leq \cdots \leq \delta^p$. Now, define $n \in \{0, \ldots, m\}$ such that $\delta^1 \leq \cdots \leq \delta^n < 0$ and

$\delta^{n+1} = \cdots = \delta^{m+1} = 0$ and construct the sets

$$\mathcal{D}^- \triangleq \mathcal{B}^1 \cup \cdots \cup \mathcal{B}^n, \quad \text{and} \quad \mathcal{D}^0 \triangleq \mathcal{B}^{n+1} \cup \cdots \cup \mathcal{B}^{m+1}, \quad \text{and} \quad \mathcal{D}^+ \triangleq \mathcal{B}^{m+2} \cup \cdots \cup \mathcal{B}^p.$$

Also recall that $\tilde{\boldsymbol{R}} \succeq \boldsymbol{R}^*$. Consequently, it follows that no user with index in $\mathcal{D}^-$ can be decoded at receiver $i$, under the rate assignment $\tilde{\boldsymbol{R}}$. Thus, the users in $\mathcal{D}^-$ must be treated as Gaussian interferers. Next since the rates of the users in $\mathcal{D}^+$ remain unaltered, these users are decodable using the partition $\{\cup_{j=m+2}^{p} \mathcal{B}^j, \mathcal{K} \setminus \cup_{j=m+2}^{p} \mathcal{B}^j\}$ under the rate assignment $\tilde{\boldsymbol{R}}$. Thus, without loss of optimality, we can assume that these users have been perfectly decoded and expurgated.

Let us focus on any arbitrary partitioning of users $\{\tilde{\mathcal{G}}, \mathcal{D}^- \cup \mathcal{D}^0 \setminus \tilde{\mathcal{G}}\}$, such that $i \in \tilde{\mathcal{G}}$. Our objective is to show that user $i$ is not decodable under the rate assignment $\tilde{\boldsymbol{R}}$ using any such partition. First consider the case $1 \notin \tilde{\mathcal{G}}$. For user $i$ to be decodable we must have $\Delta(\boldsymbol{h}^i, \tilde{\mathcal{G}}, \mathcal{D}^- \cup \mathcal{D}^0 \setminus \tilde{\mathcal{G}}, \tilde{\boldsymbol{R}}) = \Delta(\boldsymbol{h}^i, \tilde{\mathcal{G}}, \mathcal{D}^- \cup \mathcal{D}^0 \setminus \tilde{\mathcal{G}}, \boldsymbol{R}^*) \geq 0$. Using the fact that $\delta^{n+1} = 0$, we can conclude that $\Delta(\boldsymbol{h}^i, \mathcal{D}^0 \setminus \tilde{\mathcal{G}}, \mathcal{D}^-, \boldsymbol{R}^*) \geq 0$. However, since $\Delta(\boldsymbol{h}^i, \mathcal{D}^0, \mathcal{D}^-, \boldsymbol{R}^*) = 0$ we must have that both $\Delta(\boldsymbol{h}^i, \tilde{\mathcal{G}}, \mathcal{D}^- \cup \mathcal{D}^0 \setminus \tilde{\mathcal{G}}, \boldsymbol{R}^*)$ and $\Delta(\boldsymbol{h}^i, \mathcal{D}^0 \setminus \tilde{\mathcal{G}}, \mathcal{D}^-, \boldsymbol{R}^*)$ are equal to zero. Again using the fact that $\delta^{n+1} = 0$, we can conclude that $\Delta(\boldsymbol{h}^i, \mathcal{D}^0 \setminus \tilde{\mathcal{G}} \cup (\cup_{j=n+1}^{m} \mathcal{B}^j), \mathcal{D}^-, \boldsymbol{R}^*) \geq 0$. However since both $\Delta(\boldsymbol{h}^i, \cup_{j=n+1}^{m} \mathcal{B}^j, \mathcal{D}^-, \boldsymbol{R}^*)$ and $\Delta(\boldsymbol{h}^i, \mathcal{D}^0 \setminus \tilde{\mathcal{G}}, \mathcal{D}^-, \boldsymbol{R}^*)$ are zero, we must have that

$$\Delta(\boldsymbol{h}^i, (\mathcal{D}^0 \setminus \tilde{\mathcal{G}}) \setminus (\cup_{j=n+1}^{m} \mathcal{B}^j), (\cup_{j=n+1}^{m} \mathcal{B}^j) \cup \mathcal{D}^-, \boldsymbol{R}^*) = 0.$$

This yields us the desired contradiction since the set $(\mathcal{D}^0 \setminus \tilde{\mathcal{G}}) \setminus (\cup_{j=n+1}^{m} \mathcal{B}^j)$ does not contain $i$ but was not selected instead of $\mathcal{B}^{m+1}$ in step 3 of Algorithm 3. Consequently, we can conclude that $i$ is not decodable using $\{\tilde{\mathcal{G}}, \mathcal{D}^- \cup \mathcal{D}^0 \setminus \tilde{\mathcal{G}}\}$ under rate assignment $\boldsymbol{R}^*$ and hence under rate assignment $\tilde{\boldsymbol{R}}$. Finally, we need to rule out partitions $\{\tilde{\mathcal{G}}, \mathcal{D}^- \cup \mathcal{D}^0 \setminus \tilde{\mathcal{G}}\}$ such that $1, i \in \tilde{\mathcal{G}}$. For user $i$ to be decodable, we must have $\Delta(\boldsymbol{h}^i, \tilde{\mathcal{G}}, \mathcal{D}^- \cup \mathcal{D}^0 \setminus \tilde{\mathcal{G}}, \tilde{\boldsymbol{R}}) \geq 0$. Using the facts that $\delta^{n+1} = 0$ and $1 \notin \mathcal{D}^0 \setminus \tilde{\mathcal{G}}$, we can conclude that $\Delta(\boldsymbol{h}^i, \mathcal{D}^0 \setminus \tilde{\mathcal{G}}, \mathcal{D}^-, \tilde{\boldsymbol{R}}) \geq 0$. These facts together provide that $\Delta(\boldsymbol{h}^i, \mathcal{D}^0, \mathcal{D}^-, \tilde{\boldsymbol{R}}) \geq 0$. However, this is a contradiction since $\Delta(\boldsymbol{h}^i, \mathcal{D}^0, \mathcal{D}^-, \boldsymbol{R}^*) = 0$ and $\tilde{\boldsymbol{R}} \geq \boldsymbol{R}^*$ with $\tilde{R}_1 > R_1^*$.

---

**Algorithm 1** - Solving $\tilde{\mathcal{P}}(\boldsymbol{\gamma})$

1: Input $\boldsymbol{\alpha}, \boldsymbol{\gamma}, \boldsymbol{\beta}$, and $\{\boldsymbol{h}_{i,j}^{s,s}\}, \{\boldsymbol{h}_{i,j}^{s,p}\}, \{\boldsymbol{h}_{i,j}^{p,s}\}, \{\boldsymbol{h}_{i,j}^{p,p}\}$
2: Define $\{\tilde{\boldsymbol{h}}_{i,j}^{s,s}\}, \{\tilde{\boldsymbol{h}}_{i,j}^{p,s}\}$ as specified in (9)
3: Initialize $\boldsymbol{\lambda}$ and $k = 1$
4: **repeat**
5:    Construct $\boldsymbol{U}_i$ as in (12); obtain $\hat{\boldsymbol{h}}_{j,i}^{s,s} = \tilde{\boldsymbol{h}}_{j,i}^{s,s} \boldsymbol{U}_i^{-1}$
6:    Solve $g(\boldsymbol{\lambda})$ using the distributed algorithm of [7] and find $\{\hat{\boldsymbol{w}}_i^s\}$
7:    Obtain $\{\tilde{\boldsymbol{w}}_i^s\}$ using transformation $\tilde{\boldsymbol{w}}_i^s = \boldsymbol{U}_i^{-1} \hat{\boldsymbol{w}}_i^s$
8:    Calculate the subgradient $\boldsymbol{s}^{(k)}$ as in (17)
9:    Update $\boldsymbol{\lambda}^{(k+1)} = \boldsymbol{\lambda}^{(k)} - \frac{1}{k}\boldsymbol{s}^{(k)}$ and $k \leftarrow k+1$
10: **until** convergence
11: Output $\{\boldsymbol{w}_i^s\} = \{\frac{1}{\sqrt{\alpha_i}} \tilde{\boldsymbol{w}}_i^s\}$ and $\mathcal{P}(\boldsymbol{\gamma}) = \sum_i \alpha_i \|\boldsymbol{w}_i^s\|^2$

---

**Algorithm 2** - Solving $\mathcal{R}(P_0)$

1:   Input $\boldsymbol{\alpha}, \boldsymbol{\rho}, \boldsymbol{\beta}, \delta$ and $\{\boldsymbol{h}_{i,j}^{s,s}\}, \{\boldsymbol{h}_{i,j}^{s,p}\}, \{\boldsymbol{h}_{i,j}^{p,s}\}, \{\boldsymbol{h}_{i,j}^{p,p}\}$
2:   Initialize $\rho_{\min} = \min_{1 \leq i \leq M_s} \{\log(1 + \frac{\hat{\alpha} \|\boldsymbol{h}_{i,i}^{s,s}\|^2}{\hat{\alpha}\sum_{j \neq i} |\boldsymbol{h}_{i,j}^{s,s}(\boldsymbol{h}_{j,j}^{s,s})^H|^2/\|\boldsymbol{h}_{j,j}^{s,s}\|^2 + a_i^s})/\rho_i\}$ and
    $\rho_{\max} = \min_{1 \leq i \leq M_s} \{\log(1 + P_0 \frac{\|\boldsymbol{h}_{i,i}^{s,s}\|^2}{\alpha_i(a_i^s)})/\rho_i\}$
3:    $\rho_0 \leftarrow \rho_{\min}$, $\boldsymbol{\gamma} \leftarrow 2^{\rho_0 \boldsymbol{\rho}} - 1$
4:   **repeat**
5:     Solve $\tilde{\mathcal{P}}(\boldsymbol{\gamma})$ using Algorithm 1
6:     **if** $P_0 \geq \tilde{\mathcal{P}}(\boldsymbol{\gamma})$
7:       $\rho_{\min} \leftarrow \rho_0$; update $\{\boldsymbol{w}_i^s\}$
8:     **else**
9:       $\rho_{\max} \leftarrow \rho_0$
10:     **end if**
11:    $\rho_0 \leftarrow (\rho_{\min} + \rho_{\max})/2$ and $\boldsymbol{\gamma} \leftarrow 2^{\rho_0 \boldsymbol{\rho}} - 1$
12:   **until** $\rho_{\max} - \rho_{\min} \leq \delta$
13:   Output $\mathcal{R}(P_0) = \rho_{\min}$ and $\{\boldsymbol{w}_i^s\}$

**Algorithm 3** - Rate increment recommendations by individual receivers

1: Initialize $\mathcal{S} = \mathcal{K}$ and $\mathcal{G} = \emptyset$ and $\mathcal{G}^i = \emptyset$ and $k = 1, \boldsymbol{R}^{\min}$
2: **repeat**
3:     Find $\delta^k = \min_{\mathcal{B} \neq \emptyset, \mathcal{B} \subseteq \mathcal{S}} \frac{\Delta(\boldsymbol{h}^i, \mathcal{B}, \mathcal{G}, \boldsymbol{R}^{\min})}{\sum_{j \in \mathcal{B}} \rho_j}$   and   $\mathcal{B}^k = \arg\min_{\mathcal{B} \neq \emptyset, \mathcal{B} \subseteq \mathcal{S}} \frac{\Delta(\boldsymbol{h}^i, \mathcal{B}, \mathcal{G}, \boldsymbol{R}^{\min})}{\sum_{j \in \mathcal{B}} \rho_j}$
    If there are multiple choices for $\mathcal{B}^k$ pick any one such that $i \notin \mathcal{B}^k$
4:     **if** $i \in \mathcal{B}^k$ or $i \in \mathcal{G}$
5:         $r_j^i = \delta^k \rho_j$ for all $j \in \mathcal{B}^k$   and   $\mathcal{S} \leftarrow \mathcal{S} \backslash \mathcal{B}^k$ and $\mathcal{G} \leftarrow \mathcal{G} \cup \mathcal{B}^k$ and $\mathcal{G}^i \leftarrow \mathcal{B}^k \cup \mathcal{G}^i$ and $k \leftarrow k+1$
7:     **else**
8:         $r_j^i = +\infty$ for all $j \in \mathcal{B}^k$, $\mathcal{S} \leftarrow \mathcal{S} \backslash \mathcal{B}^k$ and $\mathcal{G} \leftarrow \mathcal{G} \cup \mathcal{B}^k$, $k \leftarrow k+1$
9:     **end if**
10: **until** $\mathcal{S} = \emptyset$
11: Output $\{r_k^i\}$ and $\mathcal{G}^i$

---

**Algorithm 4** - Distributed Weighted Max-Min Fair Rate Allocation

1: Initialize $\boldsymbol{R}^{\min}$ and $q = 0$
2: **repeat**
3:     **for** $i = 1, \ldots, M_s$ **do**
4:         Run Algorithm 3
5:     **end for**
6:     Update $q \leftarrow q+1$ and $R_k^{(q)} = R_k^{\min} + \min_{1 \leq i \leq M_s}\{r_k^i\}$ and $\boldsymbol{R}^{\min} \leftarrow \boldsymbol{R}^{(q)}$
7: **until** $\boldsymbol{R}^{(q)}$ converges
8: Output $\boldsymbol{R}^* = \boldsymbol{R}^{(q)}$ and $\{\mathcal{G}^i\}_{i=1}^{M_s}$

---

**Algorithm 4MLD** - Distributed Weighted Max-Min Fair Rate Allocation for MLD

1: Initialize $\boldsymbol{R}^{\min}$ and $q = 0$
2: **repeat**
3:     **for** $i = 1, \ldots, M_s$ **do**
4:         Initialize $\mathcal{S} = \mathcal{K}$
5:         **repeat**
6:             Find $\delta = \min_{\mathcal{B}: i \in \mathcal{B}, \mathcal{B} \subseteq \mathcal{S}} \frac{\Delta(\boldsymbol{h}^i, \mathcal{B}, \emptyset, \boldsymbol{R}^{\min})}{\sum_{j \in \mathcal{B}} \rho_j}$
7:             $\mathcal{B} = \arg\min_{\mathcal{B}: i \in \mathcal{B}, \mathcal{B} \subseteq \mathcal{S}} \frac{\Delta(\boldsymbol{h}^i, \mathcal{B}, \emptyset, \boldsymbol{R}^{\min})}{\sum_{j \in \mathcal{B}} \rho_j}$
8:             $r_j^i = \delta \rho_j$ for all $j \in \mathcal{B}$
9:             $\mathcal{S} \leftarrow \mathcal{S} \backslash \mathcal{B}$
10:         **until** $\mathcal{S} = \emptyset$
11:     **end for**
12:     Update $q \leftarrow q+1$ and $R_k^{(q)} = R_k^{\min} + \min_{1 \leq i \leq M_s}\{r_k^i\}$ and $\boldsymbol{R}^{\min} \leftarrow \boldsymbol{R}^{(q)}$
13: **until** $\boldsymbol{R}^{(q)}$ converges
14: Output $\boldsymbol{R}^{\mathrm{ML}} = \boldsymbol{R}^{(q)}$